\documentclass[12pt]{article}
\pdfoutput=1
\usepackage{geometry}                		
\usepackage{graphicx}				
\usepackage{amssymb}
\usepackage{multicol}
\usepackage{url}
\usepackage{color}
\usepackage{hyperref}
\usepackage{setspace}
\usepackage{graphicx}
\usepackage{stackrel}
\usepackage{mathtools}
\usepackage{amsmath}
\usepackage{mathrsfs}
\usepackage{amsfonts}
\usepackage{subfigure}
\usepackage{arydshln}
\usepackage{stackrel}
\usepackage{alltt}
\usepackage[latin1]{inputenc}
\usepackage{times}
\usepackage[T1]{fontenc}
\usepackage{pgfkeys}
\usepackage{tikz}
\usepackage{wrapfig}
\usepackage[sort]{natbib}
\usepackage{cancel}
\usepackage{booktabs}
\usepackage{pifont}
\usepackage{multirow}
\usepackage{hhline}
\usepackage{varioref}
\usepackage{caption}
\usepackage{epstopdf}
\usepackage{cancel}
\usepackage{fullpage}
\usepackage{lineno}
\usepackage{placeins}
\usepackage{setspace}

\newtheorem{theorem}{Theorem}
\newcommand{\mbf}[1]{\mathbf{#1}}
\newcommand{\mbs}[1]{\boldsymbol{#1}}

\renewcommand{\d}{\text{d}}
\newcommand{\by}{\mbf{y}}
\newcommand{\Ys}{Y^\star}

\newcommand{\yt}{\tilde{y}}
\newcommand{\Yt}{\tilde{Y}}

\newcommand{\Yst}{\tilde{Y}^\star}

\newcommand{\ysb}{\bar{y}^\star}

\newcommand{\bb}{\mbf{b}}

\newcommand{\xs}{x^\star}
\newcommand{\bz}{\mbs{\psi}}

\newcommand{\Xs}{X^\star}

\newcommand{\bL}{\mbf{L}}

\newcommand{\bv}{\mbf{v}}

\newcommand{\bQ}{\mbf{Q}}
\newcommand{\bZ}{\mbs{\Psi}}

\newcommand{\bB}{\mbf{B}}

\newcommand{\cR}{\mathcal{N}}

\newcommand{\blambda}{\mbs{\lambda}}

\newcommand{\bgamma}{{\mbs{\gamma}}}

\newcommand{\bpi}{{\mbs{\pi}}}

\newcommand{\bH}{\mbf{H}}

\newcommand{\1}{\mbs{1}}
\newcommand{\0}{\mbs{0}}

\newcommand{\MB}{{\bgamma_0}}
\newcommand{\MF}{{\bgamma_{\cR} }}

\newcommand{\cM}{\mathcal{M}}

\newcommand{\lrp}[1]{\left(#1\right)}
\newcommand{\lrb}[1]{\left\{#1\right\}}

\newcommand{\lrsqb}[1]{\left[#1\right]}

\renewcommand{\d}{\text{d}}

\newcommand{\be}{\begin{equation}}
\newcommand{\ee}{\end{equation}}
\newcommand{\ben}{\begin{equation*}}
\newcommand{\een}{\end{equation*}}
\newcommand{\bea}{\begin{eqnarray}}
\newcommand{\eea}{\end{eqnarray}}
\newcommand{\bean}{\begin{eqnarray*}}
\newcommand{\eean}{\end{eqnarray*}}
\newcommand{\bsmat}{\begin{smallmatrix}}
\newcommand{\esmat}{\end{smallmatrix}}
\newcommand{\bmat}{\begin{matrix}}
\newcommand{\emat}{\end{matrix}}

\title{On the estimation of the order of smoothness of the regression function\footnote{This material was based upon work partially supported by the National Science Foundation under Grant DMS-1127914 to the Statistical and Applied Mathematical Sciences Institute. Any opinions, findings, and conclusions or recommendations expressed in this material are those of the author(s) and do not necessarily reflect the views of the National Science Foundation.}}
\author{Daniel Taylor-Rodriguez \thanks{Postdoctoral Fellow, Duke University / SAMSI, RTP, NC 27709. Email:  \texttt{taylor-rodriguez@samsi.info}}\\
Sujit  K. Ghosh \thanks{Professor, NCSU / Deputy Director, SAMSI, RTP, NC 27709. Email:  \texttt{ghosh@samsi.info}}}
\date{\today} 

\begin{document}

\maketitle

\begin{abstract}
The order of smoothness chosen in nonparametric estimation problems is critical.  This choice balances the tradeoff between model parsimony and data overfitting.  The most common approach used in this context is cross-validation.  However, cross-validation is computationally time consuming and often precludes valid post-selection inference without further considerations. With this in mind, borrowing elements from the objective Bayesian variable selection literature, we propose an approach to select the degree of a polynomial basis.  Although the method can be extended to most series-based smoothers, we focus on estimates arising from Bernstein polynomials for the regression function, using mixtures of g-priors on the model parameter space and a hierarchical specification for the priors on the order of smoothness.  We prove the asymptotic predictive optimality for the method, and through simulation experiments, demonstrate that, compared to cross-validation, our approach is one or two orders of magnitude faster and yields comparable predictive accuracy.  Moreover, our method provides simultaneous quantification of model uncertainty and parameter estimates. We illustrate the method with real applications for continuous and binary responses.
\end{abstract}

\begin{spacing}{1}

\section{Introduction}

Consider the following regression problem. Let $\lrp{X,Y}\in\lrsqb{0,1}\times\mathbb{R}$ be a random vector such that $E\lrsqb{Y^2}<\infty$ and $E\lrsqb{Y|X=x}\equiv\mu(x)$, where $\mu(\cdot)$ is continuous on $\lrsqb{0,1}$. For a random sample of $n$  observations $\lrb{\lrp{X_1,Y_1},\ldots,\lrp{X_n,Y_n}}$, we assume the model is given by
\bea
Y_i &=& \mu(X_i)+\sigma\epsilon_i, \text{ for }i=1,\ldots,n, 
\nonumber
\eea
where $\sigma>0$ and $\mu(\cdot)$ are parameters to be estimated,  and $\epsilon$ is such that $E\lrsqb{\epsilon|x}=0$ and $\text{Var}\lrp{\epsilon|x}\leq1\;\forall x\in\lrsqb{0,1}$. 

Establishing the functional relationship between the response $Y$ and the predictor $X$ has been a recurring theme in the statistical and mathematical literature for the past several decades.  Parametric forms determine a priori the functional form of this relationship. As such, these are limited in their ability to uncover the true nature of the association.  On the other hand, a plethora of nonparametric methods \citep[see][and references therein]{Ruppert2003}, which adapt to nonlinear features present in the data and enjoy good theoretical properties, have been extensively studied \citep[e.g.,][]{Stadtmuller1986}.  

Nonparametric methods can be classified roughly into four categories: smoothing splines \citep{Eubank1985, Craven1978,Silverman1985}, regression splines \citep{Friedman1991, Wang2002, Wood2003}, penalized splines \citep{Eilers1996,Eilers2010}, and kernel-based methods \citep{Nadaraya1964, Ramsay1991}.   The efficacy of nonparametric techniques in approximating $\mu(\cdot)$ is highly dependent on the choice of the {\sl smoothness parameters} that regulate the balance between model generality and fidelity to the observed data.  In principle, the nonparametric take on the problem described above rests on the fact that an unknown square integrable function $\mu(\cdot)$ can be represented using basis function expansions of the form
\bea
\mu(x)&=& \sum_{k=0}^\infty \beta_k \psi_k(x),\label{eq:nonparreg}
\eea
where the known functions $\psi_0(\cdot),\psi_1(\cdot),\psi_2(\cdot),\ldots$ form a basis with respect to a suitably chosen metric norm. It is well-known a fact that any square integrable function can be approximated arbitrarily well by a smooth function supported on a compact set \citep[e.g., see Appendix A in][]{Gyorfi2002}.  

However, in practice, having an infinite or an excessively large number of terms in the sum leads to overfitting the model to the observed data.  To prevent this, the smoothness of the function is controlled by selecting a suitable number of terms; we refer to this number of terms as the {\sl order of smoothness}.  Our goal is then to estimate a smooth function by determining where to truncate the infinite series, and provide valid inference about the order itself.

In this context, cross-validation is one of the most popular procedures used to select the order of smoothness \citep[see][]{Bouezmarni2007,Leblanc2010}. In spite of this, cross-validation is computationally costly (time-wise), and the inference conducted on the model parameters after cross-validation \citep[and all other available methods used to select the order of smoothness, e.g.,][]{Mante2015}, commonly ignores additional uncertainty arising from the selection procedure \citep[see][for a general description of this issue]{Berk2012}.  However, a procedure to  choose optimally and automatically the degree of the polynomial while accounting for uncertainty, has remained elusive. 

As an alternative, in this article we introduce an approach that finds a suitable order of smoothness (hereafter denoted by $\cR^\star$), by approximately minimizing a predictive loss, while assessing the uncertainty associated with this choice.  The proposed method is automatic, in the sense that no user input is required (as ``noninformative'' priors are used on model parameters), it is computationally fast, and, can be extended to the case with multiple predictors and discrete-valued responses.  As shown in Section 5, our method yields results equivalent to those obtained with cross-validation in terms of accuracy, while taking only a fraction of the time and selecting more parsimonious models than cross-validation.

 Among the many nonparametric techniques available, our interest lies in the class of series-based smoothers, focusing on approximations to the unknown regression function $\mu(\cdot)$ with Bernstein polynomials (BPs).  This choice facilitates the exposition of the procedure and benefits from the long list of virtues that BPs possess \citep[see][ and references therein]{Farouki2012}.  

It is well known that Bernstein bases span the space of continuous functions, $C\lrsqb{0,1}$, but the elements of this basis are not orthogonal. Orthogonality of the basis is desirable as it guarantees {\it permanence} of the coefficients with respect to the degree of the approximation; meaning that, in approximations with orthogonal polynomials of orders $K$ and $K^\prime$ (for $K\not=K^\prime$ and $k\leq\min{\lrb{K,K^\prime}}$), the order-$k$ coefficients  coincide.   Selection of the order of smoothness cannot be performed directly on the Bernstein form; however, linear maps can be built between Bernstein polynomials and other orthogonal bases  \citep[see][and references therein]{Lee2002, Farouki2012}.  Moreover, it has been shown under orthogonality, using  squared error loss, that the Bayesian median probability model (MPM) is optimal for prediction, provided that some additional conditions hold \citep[see][]{Barbieri2004}. In Section 2, we provide some more details about BPs and their connection to other polynomial basis.

The existence of linear maps between Bernstein and orthogonal bases implies that choosing the order of smoothness in the space of orthogonal polynomials imposes an order onto the Bernstein polynomial. Hence, since each choice for the order of smoothness is associated to a specific model, we may cast this as a Bayesian variable selection problem (with an orthogonal base) where the polynomial hierarchy of term inclusion is respected \citep{Chipman1996,Taylor-Rodriguez2015}.  This type of constrained selection implies that, for the $k$th degree basis to be included in the model, the inclusion of all coefficients of order $k-1$ and lower is necessary. Conversely, if the $k$th degree basis is excluded from the model, this constraint prevents any term of order greater than $k$ from being included.  


The article is organized as follows.   Section 2 provides details about BPs, emphasizing the connection to other orthogonal polynomial bases that will allow us to make use of variable selection tools. In Section 3 we introduce some underlying concepts on Bayesian selection, formulate the problem and elaborate on the proposed methodology. Section 4 provides some results demonstrating, under squared error loss, the asymptotic predictive optimality for the order of smoothness selected. In Section 5, through simulation experiments we assess the performance of our approach and compare it to that of cross-validation. Section 6 illustrates the use of the methodology with two case studies and describes a simple adaptation to incorporate binary responses. Finally, we close by providing some concluding remarks.

\section{Preliminaries: Bernstein polynomials and orthogonal bases}

Bernstein polynomial type estimators of regression functions were first developed by \citet{Stadtmuller1986} for the single predictor case. \citet{Tenbusch1997} extended the methodology for the multiple predictor regression problem, and demonstrated that Bernstein polynomial type estimators are pointwise and uniformly consistent, and are asymptotically normal.  These properties are also inherited by its derivatives.  In addition to the above, the Bernstein basis is the only {\it optimally-stable} basis function in use \citep{Farouki2012}.  This implies that, among all non-negative polynomial bases, its coefficients are the least vulnerable to the effects of random perturbations. Theoretical underpinnings aside, curve estimation with BPs is straightforward, can be used with censored data \citep{Osman2012}, and allows imposing shape constraints on the curve effortlessly whenever background information to do so is available  \citep{McKayCurtis2011, Wang2012a}.  

The BP approximation of degree $\cR\in\mathbb{N}\cup\lrb{0}$ to the function $\mu(\cdot)$ at a value $x\in\lrsqb{0,1}$ is given by a sequence of polynomials of the form
\bea
\mu^{(\bB)}_\cR(x)&=&\sum_{k=0}^\cR\eta_k^{\cR} b_k^{\cR}(x) \nonumber\\
&=& \lrp{b_0^\cR(x), \bb_\cR(x)^\prime}^\prime{\eta_0 \choose \mbs{\eta}_\cR},
\label{eq:BernApprox}
\eea  
where one may choose $\eta_0^{\cR}=\mu\lrp{0}$ and $b_0^{\cR}(x) =(1-x)^{\cR}$, and $\eta_k^{\cR}=\mu\lrp{k/\cR}$ and $b_k^{\cR}(x) ={\cR\choose k} x^k (1-x)^{(\cR-k)}$, for $k=1,\ldots,\cR$.  Other choices for $\eta_k^\cR$ can be obtained using through iterated Bernstein polynomial operators \citep[see][]{Kelisky1967,Mante2015}. 

The choice of $\cR$ is critical in producing a good approximation as it is well know that, if $\mu(\cdot)$ satisfies an order $\alpha\in(0,1]$ Lipschitz condition, then \citep{Berens1972}
\[\|\mu(\cdot)-\mu^{(\bB)}_\cR(\cdot)\|_\infty \equiv \underset{0\leq x\leq 1}{\sup} |\mu(x)-\mu^{(\bB)}_\cR(x) |=O(\cR^{-\alpha/2}).\]   
Asymptotically, \citet{Stadtmuller1986} obtained as an optimal choice $\cR=O(n^{2/5})$; similarly \citet{Tenbusch1997} suggested using some integer $\cR\in \lrsqb{n^{2/5},n^{2/3}}$  (provided certain conditions hold, see \citet{Tenbusch1997} for details).  

Having these values to guide the selection of the order of the polynomial is useful, but once the sample size becomes moderately large, having the order of smoothness lie in $\lrsqb{n^{2/5},n^{2/3}}$ can make the method susceptible to overfitting (see simulations in the Appendix). As such, an integer in the interval $\lrsqb{n^{2/5},n^{2/3}}$ could instead be taken as an upper bound for the order of the Bernstein polynomial rather than being used as the order of the polynomial itself.  Throughout the remainder of the article, we will take $\cR=\lfloor n^{2/3}\rfloor$ as the upper bound on the order of smoothness.

In principle, the Bernstein degree-$\cR$ approximation given in \eqref{eq:BernApprox} may be represented in terms of  any degree-$\cR$ polynomial basis. Any such change of basis is obtained through a linear map, which produces the coefficients of the new basis by multiplying the coefficients from the original one by a transformation matrix $\bQ_\cR$ \citep{Farouki2012}.  

The linear map between the coefficients of the Bernstein basis and those of an orthogonal basis (characterized by a transformation matrix $\bQ_\cR$) is extremely convenient for the problem at hand.  One may first select the order of smoothness by choosing the polynomial model of order $\cR^\star$ with the orthogonal basis, estimate its parameters, and finally, transform these estimates into their BP form using the transformation matrix $\bQ_{\cR^\star}$.  Alternatively, to avoid the loss of stability that arises from going from one basis to another, one may simply use the selected order of smoothness $\cR^\star$ under the orthogonal representation, and fit directly the order-$\cR^\star$ BP to the data to obtain the parameter estimates. 

\subsection{Choice of orthogonal basis}

The Bernstein basis is the only {\it optimally-stable} basis function in use \citep{Farouki2012}. Optimal stability implies that, among all non-negative polynomial-bases, its coefficients are least vulnerable to random perturbations. Consequently, it is usually recommended to avoid such transformations to prevent increased amplification of random perturbations in the coefficients.  Transformation stability from the Bernstein to  other bases has been studied extensively \citep[see][and references therein]{Farouki2012}, and is commonly assessed with what is referred to as the {\it condition number}. The {\it condition number} quantifies how much a random perturbation on an input (the coefficients) impacts the output (the curve estimate).  The choice of orthogonal basis is all-important to control the loss of stability in the estimates.  Among the transformations between Bernstein and other commonly used bases (e.g., power, Chebychev and Legendre), that onto Legendre polynomials has been shown to retain the most stability \citep{Farouki2000}. 

The Legendre polynomial of degree $\cR$ in $\lrsqb{0,1}$ is defined by the recurrence relation defined by
\bean
\psi_0(x)&=&1\\
\psi_1(x)&=&2x-1\\
(k+1)\psi_{k+1}(x)&=&(2k+1)(2x-1)\psi_k(x)-k\, \psi_{k-1}(x),\quad k=1,2,\ldots
\eean
The order $\cR$ approximant  to $\mu(x)$ in Legendre form is given by \citep{Farouki2012}
\bea
\mu^{(\bL)}_\cR(x) &=& \sum_{k=0}^{\cR} \lambda_k \psi_k(x)\nonumber\\
&=&\lrp{\psi_0(x), \mbs{\psi}_\cR(x)^\prime }^\prime {\lambda_0 \choose \blambda_\cR}
\label{eq:LegApprox}
\eea
In addition to the recursive expression for the orthogonal basis $\psi_k(x)$, a more elegant characterization for Legendre polynomials  in terms of Bernstein basis functions is
\[\psi_k(x)=\sum_{j=0}^{k} (-1)^{k+j} {k\choose j} b_j^k(x). \]

\citet{Farouki2000} studied transformations between the Legendre and Bernstein order-$\cR$ polynomials defined on $\lrsqb{0,1}$, deriving explicit forms for the elements that relate $\eta_0^\cR,\cdots,\eta_\cR^\cR$ and $\lambda_0,\cdots,\lambda_\cR$ in \eqref{eq:BernApprox} and  \eqref{eq:LegApprox} respectively, through the linear transformation $\eta_j^\cR=\sum_{k=0}^\cR Q_{jk}^{(\cR)} \lambda_k$. More generally ${\eta_0 \choose \mbs{\eta}_\cR}=\bQ_{\cR} {\lambda_0 \choose \blambda_\cR}$, where the coefficients $Q_{jk}^{(\cR)}$ of the transformation matrix $\bQ_{\cR}$ are defined as
 \bea
 Q_{jk}^{(\cR)}&=& \frac{1}{{\cR \choose j}} \sum_{i=\max{(0,j+k-\cR)}}^{\min{(j,k)}}(-1)^{k+i} \lrsqb{{k \choose i}}^2{\cR-k \choose j-i},
 \label{eq:Qcoeff}
 \eea
for $0\leq j,k\leq \cR$.  Conversely, the Legendre coefficients may be derived from the Bernstein  coefficients with the reverse mapping $ \blambda_\cR=\bQ_{\cR}^{-1}\mbs{\eta}_\cR$, where the elements of $\bQ_{\cR}^{-1}$ are
 \bea
 \lrp{Q_{jk}^{(\cR)}}^{-1}&=&\frac{2j+1}{\cR+j+1}{{\cR \choose k}} \sum_{i=0}^{j}(-1)^{j+i} \frac{{j \choose i}{j \choose i}}{{\cR+j \choose k+i}}
 \label{eq:Qinvcoeff}
 \eea
The expressions in \eqref{eq:Qcoeff} and \eqref{eq:Qinvcoeff} can also be computed recursively, avoiding the need to compute the combinatorial terms that may render the calculations numerically unstable.

Using the linear association between Bernstein and Legendre basis described in this section one may identify the order of smoothness for the Bernstein polynomial.  In Section 3 we propose an alternative method, which consists of formulating a selection on a polynomial model space that obeys heredity constraints.

\section{Methodology}

The goal is to formulate a procedure with good operating properties to determine automatically a parsimonious (data-dependent) Bernstein polynomial approximation for $\mu(x)$.  Suppose that for a random value of the predictor $\Xs\sim X$ we want to predict a value from $\Ys=\mu(\Xs)+\sigma \epsilon^\star$, where $\mu(\cdot)$ is the true underlying value for the regression function and $\epsilon^\star\sim\epsilon$.  Now, for a given $k\in\lrb{0,1,\ldots,\cR}$, note that $\Ys$ under the order-$k$ Bernstein polynomial model is given by $Y_k^\star=\mu_k^{(\bB)}(\Xs)+\sigma\epsilon^\star$, where $\mu_k^{(\bB)}(\cdot)$ is defined as in \eqref{eq:BernApprox}, and denote by $\Yt_k^\star=\mathbb{E}\lrsqb{Y_k^\star \,|\,(x_1,y_1),\ldots,(x_n,y_n)}$.   The order of the optimal approximation $\cR^*\in\lrb{0,1,\ldots,\cR}$ (for $\cR=\lfloor n^{2/3}\rfloor$) is defined as
\bea
\cR^\star&=&\underset{k\in\lrb{0,\ldots,\cR}}{\text{argmin}}\mathbb{E}\lrsqb{\lrp{\Ys-\Yst_k}^2 \Big|\, (x_1,y_1),\ldots, (x_n,y_n)},\label{eq:Rstar}
\eea
where the posterior expectation is with respect to the posterior distribution of $(\Xs,\Ys)$ given the data $(x_1,y_1),\ldots,(x_n,y_n)$.  In other words, $\cR^\star$ is the degree of the polynomial that minimizes the posterior predictive squared error loss.

To obtain $\cR^\star$, we formulate the nonparametric estimation problem with BPs in a way that makes tools from the objective Bayesian testing literature compatible with this problem. As discussed in Section 2, it is first necessary to move onto the Legendre basis, which is orthogonal, and preserves, as much as possible, the desirable features of the Bernstein form. In the Legendre space, once the parameter and model prior probabilities have been defined, the variable selection procedure can be implemented, provided a method for estimating the posterior probabilities is available.  In the case of a polynomial basis generated from a single predictor, the size of the model space corresponds to the number of terms in the highest order model considered ($\cR+1$ for the problem at hand).  Since the model space can be easily enumerated, model posterior probabilities can be calculated in closed form, precluding the need for a stochastic search algorithm to explore the model space. For problems where two or more predictors are considered, and a multivariate Bernstein polynomial is used for nonparametric estimation, the stochastic search algorithm proposed in \citet{Taylor-Rodriguez2015} may be used.

In the remainder of this section, we elaborate on how the Legendre representation enables choosing the order of smoothness via a Bayesian selection algorithm that respects the polynomial structure in the predictor space of the nonparametric models being considered.

To begin with, let model space $\cM$ (in its Legendre form) correspond to the set of models $\lrb{\MB,\bgamma_1,\ldots,\bgamma_\cR}$, where the  order-$k$ polynomial model is given by $\bgamma_k=(\gamma_{0},\gamma_{k,1},\gamma_{k,2},\ldots,\gamma_{k,\cR})$, where $\gamma_{k,j}=1$ if $j\leq k$ and  $\gamma_{k,j}=0$ for $j>k$.  That is, the $\gamma_{k,j}$ is the inclusion indicator variable for the order-$j$ term in $\bgamma_k$. Hereinafter, we assume that all probabilistic statements are conditional on model space $\cM$ and for simplicity avoid conditioning on it explicitly.

In the Bayesian paradigm model $\bgamma_k\in\cM$ is characterized by a sampling density and a prior distribution.   The sampling density associated with $\bgamma_k$ is $f(\by |\lambda_0,\blambda_{k},\tau,\bgamma)$, where $\lambda_0,\blambda_{k}$, and $\tau$ are the parameter in the base model $\MB$, the vector of parameters included in $\bgamma_k$ different from $\lambda_0$, and the precision parameter, respectively.  The prior probability for model $\bgamma_k$ and its corresponding set of parameters is $\pi (\lambda_0,\blambda_k,\tau,\bgamma_k)=\pi (\lambda_0,\blambda_{k},\tau)\, \pi (\bgamma_k)$.  In this section we introduce the priors used on the parameters and the models.

\subsection{Parameter priors: Mixtures of $g$-priors with normal response}

Objective local priors for the model parameters $(\lambda_0,\blambda_{k},\tau)$ are achieved through modifications and extensions of Zellner's $g$-prior \citep{Berger1996,Moreno1998,Liang2008,Womack2014}.  These are referred to as scaled mixtures of $g$-priors, and in general, they share similar good operating properties with slight differences in their behavior near the origin. The distribution assumed for the precision parameter ($\omega$ in the equations below) determines the type of mixture of $g$-prior being considered; the possible priors considered for this parameter are described in Section 3.3.

Let matrices $\bZ_0$ and $\bZ_k$ contain the Legendre bases (evaluated at the observed values for $X$) related to parameters $\lambda_0$ and $\blambda_k$, respectively. The parameters $\lambda_0$ and $\tau$ that conform the base model $\MB$,  are assigned a reference prior (e.g., Jeffreys prior) and are assumed to be within every model in $\cM$.  The likelihood function is given by $\by|\blambda,\bgamma_k\sim \mathcal{N}\lrp{\lambda_0\bZ_0+\bZ_k \blambda_k ,\tau \mbf{I}}$; letting $q_k=k+1$ (i.e., the number of regression coefficients in $\bgamma_k$), the form of the scaled mixtures of $g$-priors is
\bea
\pi(\lambda_0,\blambda_k,\tau|\omega)&=& c_0 \, \tau\times
\mathcal{N}_{q_k}\lrp{\blambda_k\quad\Bigg|\quad \0,\lrp{ \frac{\omega\, \tau (q_k+1)}{n}\bZ_{k}^\prime\left(\mbf{I}-\mbf{P}_0\right)\bZ_{k}}^{-1}}
\label{eq:gpriors}
\eea
where $g=1/\omega$ in the $g$-prior formulation, $\mbf{P}_0$ is the hat matrix for $\bZ_{0}$.  Given the orthogonality of the Legendre polynomial, $\pi(\blambda_0,\blambda_k,\tau|\omega)$ further reduces to
\bea
\pi(\lambda_0,\blambda_k,\tau|\omega)&=& c_0 \, \tau\times
\mathcal{N}_{q_k}\lrp{\blambda_{k}\,\big|\, \0,\lrp{\frac{\omega\, \tau (q_k+1)}{n}\bZ_k^\prime\bZ_k}^{-1}},
\label{eq:gprortho}
\eea
with $\bZ_k^\prime\bZ_k=\text{diag}\lrb{\tilde{\bz}_j^\prime\tilde{\bz}_j:\gamma_j\in\bgamma\setminus\bgamma_0}$, where $\tilde{\bz}_j$ denotes the $j$th column of $\bZ_k$, for $n$ large and the predictor values in their original scale (i.e., $x_1,\ldots,x_n$) are uniformly distributed. Although these assumptions might not strictly hold in practice, having a moderate to large sample size suffices for $\bZ_k^\prime\bZ_k$ to be approximately diagonal.

\subsection{Priors on the model space}

Different approaches have been developed to account for structure in the predictor space within variable selection procedures  \citep[e.g.,][]{Bien2013,Yuan2009,Chipman1996}.  Among them, the methodology proposed by \citet{Chipman1996} translates beliefs about model plausibility into prior distributions, thus, helping to account for the hierarchical structure among polynomial predictors. 

The type of heredity constraints in model spaces with predictors that have a polynomial hierarchical structure can be translated into prior probability functions by relying on the inheritance principle. The inheritance principle assumes that the inclusion of a higher order term only depends on the inclusion indicators of lower-order terms from which it inherits ({\it e.g.,} the inclusion of $x_1^3$ only depends on the inclusion of $x_1$ and $x_1^2$). The construction for the model prior probabilities can be further simplified by assuming {\sl immediate inheritance}. This principle states that the inclusion of a given predictor (say $x_1^3$) conditional on the inclusion of its immediate ancestors ($x_1^2$ in this case), is independent from the inclusion of any other term in the model space.

As mentioned before, each model $\bgamma_k$ is associated to a binary vector $\bgamma_k=\lrp{1,\gamma_{k,1},\ldots,\gamma_{k,\cR}}$, with $\gamma_{k,j}= \text{I}(j\leq k)$, 
representing the inclusion status for the $\cR+1$ polynomial terms, and the order 0 term, which is always included, in the order-$k$ model.  For the nonparameric problem in a single predictor, assuming conditional independence and immediate inheritance, and, following the approach of \citet{Taylor-Rodriguez2015}, the model prior probabilities are
\bea
\pi(\bgamma_k)&=&\int \lrp{\prod_{j=1}^\cR \pi_{k,j}^{\gamma_{k,j}} (1-\pi_{k,j})^{1-\gamma_{k,j}}}p(\bpi)d\bpi
\eea
wit $\gamma_{k,j}$ defined as above, $\pi_{k,j}=\Pr(\gamma_{k,j}=1| \gamma_{k,j-1})$, and $p(\mbs{\pi})=\prod_{k=1}^\cR p_\pi(\pi_k)$ is the prior on the inclusion probabilities for each term, with

\[p_\pi(\pi_{k,j})=\left\{\begin{matrix} \text{Beta}\lrp{a_{k,j},b_{k,j}}&\text{if }\gamma_{k,j-1}=1\\ 0&\text{if }\gamma_{k,j-1}=0 \end{matrix}\right. \text{ for }k=1,\ldots,\cR.\]

The prior density $p_\pi(\pi_{k,j})$ reflects the fact that whenever the order $j-1$ term is excluded, terms of orders $j$ and higher should also be excluded. As such, the inclusion probabilities $\pi_{k,j}$ only enable paths to models that respect the polynomial hierarchy.

\subsection{Precision parameter and model posterior probabilities}

The prior specification is completed by assigning a distribution to $\omega$. Considering different priors on $\omega$ yields different objective priors.  In particular, the intrinsic prior assumes that $\omega\sim\text{Beta}(\frac{1}{2},\frac{1}{2})$.  Also a version of the Zellner-Siow prior is given by $\omega\sim \text{Gamma}(\frac{\nu}{2},\frac{\rho}{2})$ parameterized to have mean $\frac{\nu}{\rho}$, which produces a multivariate Cauchy distribution on $\blambda$ when $\nu=\rho=1$.  And finally, a family of hyper-$g$ priors is given by $\omega|\rho\sim \text{Gamma}(\frac{\nu}{2},\frac{\rho}{2})$, where $\rho\sim\text{Gamma}(\frac{a}{2},\frac{b}{2})$, with $\nu=1,a=2$, and $b=1$ recommended.  These latter ones have Cauchy-like tails but produce more shrinkage than the Cauchy prior \citep{Womack2014}.

Assuming the priors described above for the parameters and the models, the Bayes factor for $\bgamma_k$ relative to $\bgamma_0$ is given by
\[
BF_{\bgamma_k,\bgamma_0}(\mbf{y})=\left(1-R_k^2\right)^{\frac{n-q_{0}}{2}}\int\left(\frac{n+\omega(q_k+1)}{n+\frac{\omega(q_k+1)}{1-R_k^2}}\right)^{\frac{n-q_k}{2}}
\left(\frac{\omega(q_k+1)}{n+\frac{\omega( q_k+1)}{1-R_k^2}}\right)^{\frac{q_k-q_{0}}{2}}\pi(\omega)\d \omega,
\]
with $R_k^2=\by^\prime\bZ_k(\bZ_k^\prime\bZ_k)^{-1}\bZ_k^\prime \by/\by^\prime(\mbf{I}-\mbf{P}_0)\by$.

The model posterior probabilities, conditional on the observed responses $\by$ for a model $\bgamma_k\in\cM$ can be obtained in terms of its Bayes factor with respect to the base model $\bgamma_0$ as
\bea
p(\bgamma_k | \by) &=&  \frac{BF_{\bgamma_k,\bgamma_0}(\by)\pi(\bgamma_k)}{\sum_{\bgamma\in \cM} BF_{\bgamma,\bgamma_0}(\by) \pi(\bgamma)},\label{eq:postBF}
\eea

\section{Optimality under the Median Probability Model}

It is well known that the Bayesian model averaging estimators are optimal for prediction under squared error loss \citep{Raftery1997}.  However, if a single model is of interest among a particular set of models, \citet{Barbieri2004} demonstrated that in many scenarios the MPM is optimal for prediction under squared error loss.  Their results rest on the assumption that the posterior mean for the parameters of any model inside model space $\cM$, simply corresponds to the relevant coordinates from the posterior mean for the full model. Unfortunately, this assumption does not hold when using mixtures of g-priors, hence, the results obtained in \citet{Barbieri2004} are not directly applicable. That being said, for the problem of interest, an analogous result can be ascertained asymptotically, as shown below.

First, let $\bar{\lambda}_0$ and $\bar{\blambda}$ represent the resulting model averaging parameter estimates for the parameters in the full model $\MF$.  Now, denote the least squares estimate for $\blambda_\cR$ by $\hat{\blambda}_\cR=(\bZ_\cR^\prime\bZ_\cR)^{-1}\bZ_\cR^\prime \by$, and by $\hat{\blambda}_k=(\bZ_k^\prime\bZ_k)^{-1}\bZ_k^\prime \by$, that for $\blambda_k$.  Additionally, let $\tilde{\blambda}_k$ be the posterior expectation of $\blambda_k$ under model $\bgamma_k$. With mixtures of g-priors, this posterior expectation is given by $\tilde{\blambda}_k=\xi_k \hat{\blambda}_k$, where  
$\xi_k=\mathbb{E}_\omega\lrsqb{\frac{n}{n+\omega (q_k+1)}|\by,\bgamma_k}$ is the shrinkage for $\blambda_k$  \citep{Womack2014}.  

Given $\xs\in\lrsqb{0,1}$, defining $\Ys=Y(\xs)$, and conditioning on model space $\cM$, the model averaging prediction for a new response is given by $\ysb=\mathbb{E}_{\Ys|\by}\lrsqb{\,\Ys|\by}$, which depends on the $\cR$-vector of Legendre basis functions of orders 1 and higher, denoted by $\bz_\cR^\star=\bz_\cR(\xs)$.   Finally, let $\bH_k$ be the $\cR\times k$ matrix with elements $h_{jj}=1$ and $h_{mj}=0$, where $m\not=j$, $j=1,\ldots,k$ and $m=1,\ldots,\cR$. Matrix $\bH_k$ identifies the active predictors in model $\bgamma_k$, such that $\hat{\blambda}_k=\bH_k^\prime \hat{\blambda}_\cR$ and $(\bz_\cR^\star)^\prime \bH_k=(\bz_k^\star)^\prime$. The model averaging predictor is then given by
\bea
\ysb &=& \bar{\lambda}_0+(\bz_\cR^\star)^\prime \bar{\blambda}\;=\;\hat{\lambda}_0+(\bz_\cR^\star)^\prime \lrp{\sum_{k=1}^{\cR}p(\bgamma_k|\by) \bH_k \tilde{\blambda}_k}\nonumber\\
&=& \hat{\lambda}_0+(\bz_\cR^\star)^\prime \lrp{\sum_{k=1}^{\cR}p(\bgamma_k|\by) \bH_k \lrp{ \xi_k \hat{\blambda}_k}}\nonumber\\
&=& \hat{\lambda}_0+(\bz_\cR^\star)^\prime \lrp{\sum_{k=1}^{\cR}\xi_k p(\bgamma_k|\by) \bH_k \bH_k^\prime }\hat{\blambda}_\cR,
\eea
where $p(\bgamma_k|\by)$ is the posterior probability of some model $\bgamma_k\in\cM$.  

Now, define $p_j=\sum_{k=1}^{\cR} p(\bgamma_k|\by)\gamma_{k,j}$ and $\tilde{p}_j=\sum_{k=1}^{\cR}\xi_k p(\bgamma_k|\by)\gamma_{k,j}$ with $j=1,\ldots,\cR$. Recall that $\cR$ is the number of terms in $\MF$ excluding the order 0 term. Lastly, let $\yt_k^\star=\tilde{\mu}_k (\xs)$, which is the posterior predictive mean conditional on $\Xs=\xs$ under model $\bgamma_k$.  Hence, given the model space $\cM$, the posterior predictive loss with respect to model $\bgamma_k$ for a new response $\Ys=Y(\xs)$, is given by
\bea
\mbs{L}_\cR(\bgamma_k,\xs)&=& \mathbb{E}_{\Ys |\by}\lrsqb{(\Ys-\yt_k^\star )^2 | \,\by} 
\;=\; \mathbb{E}_{\Ys |\by}\lrsqb{(\Ys-\ysb+\ysb-\yt_k^\star )^2 | \,\by} \nonumber\\
&=&\text{Var}(\Ys|\by)+(\ysb-\yt_k^\star)^2 \nonumber\\
&=& \text{Var}(\Ys|\by) + \sum_{j=1}^{\cR}\lrp{\hat{\lambda}_j d_j}^2\lrp{\tilde{p}_j-\xi_k \gamma_{k,j}}^2 , \label{eq:loss}
\eea
where $d_j$ represents the $j$th diagonal element of $\bz^\star \lrp{\bz^\star}^\prime$, which for large $n$ can be approximately equal to $\tilde{\bz}_j^\prime\tilde{\bz}_j$, the $j$th diagonal element of $\bZ^\prime\bZ$. The optimal model is the one that minimizes $\mathbb{E}\lrsqb{\mbs{L}_\cR(\bgamma_k,\Xs)}$, where the expectation is over $\Xs\sim X$.  Note, that the choice of $\cR^\star$ only depends on $\by$, and that this minimization amounts to minimizing the second term on the right hand side of \eqref{eq:loss}, as $\text{Var}(\Ys|\by)$ is not a function of $\bgamma_k$.  Below we demonstrate that when using an orthogonal basis in conjunction with mixtures of g-priors for the nonparametric regression problem, this minimization is asymptotically attained by the MPM.
\begin{theorem}
Assuming a mixture of $g$-priors for the model parameters, the order chosen by the median probability model is asymptotically optimal for prediction, under the loss function in \eqref{eq:loss}, if $\frac{\cR}{n}\rightarrow 0$ as $n\rightarrow\infty$ and $\mathbb{E}\lrsqb{\omega|\by,\MF}\overset{a.s.} {\rightarrow}\omega_0\in[0,\infty)$.
\end{theorem}

\paragraph{Proof}  For every model in $\cM$ and any observed response vector $\by$, the shrinkage $\xi_k$ is bounded above by $1$. Additionally, since the shrinkage is decreasing in the number of predictors then $\xi_{\cR}\leq\ldots\leq\xi_{1}\leq\xi_{0}\leq1$.  By Jensen's inequality, we have that the shrinkage for the full model is such that
\bean
\xi_\cR\quad=\quad\mathbb{E}\lrsqb{\frac{n}{n+\omega (\cR+2)} \Big|\by,\MF}
&\geq&\frac{1}{1+\frac{\cR+2}{n}\mathbb{E}\lrsqb{\omega|\by, \MF} }
\eean
where the expression on the right hand side of the inequality tends to 1 a.s. as $n\rightarrow\infty$ if $\cR/n\rightarrow 0$ and $\mathbb{E}\lrsqb{\omega|\by, \MF}\rightarrow \omega_0\in[0,\infty)$.   Hence, $\xi_k\overset{a.s.} {\rightarrow} 1$ for all $k=0,1,\ldots,\cR$.

Let $\mbs{L}_\cR^0(\bgamma_k,\xs)= \text{Var}(\Ys|\by)+\sum_{j=1}^\cR \lrp{\hat{\lambda}_j d_j}^2\lrp{p_j-\gamma_{k,j}}^2$, which is minimized at the MPM for all $n$.  Now, consider the difference
\bean
\mbs{L}_\cR(\bgamma_k,\xs)-\mbs{L}_\cR^0(\bgamma_k,\xs) &=& \sum_{j=1}^\cR \lrp{\hat{\lambda}_j d_j}^2\lrsqb{\lrp{\tilde{p}_j-\xi_k \gamma_{k,j}}^2-\lrp{p_j-\gamma_{k,j}}^2 },
\eean
which tends to 0 a.s. for all $\xs$ as $n\rightarrow\infty$, since $\xi_k\overset{a.s.} {\rightarrow} 1$ for all $\bgamma\in\cM$.  Hence, the loss function in \eqref{eq:loss} is asymptotically equivalent to $\mbs{L}_\cR^0(\bgamma_k,\xs)$ for each $\xs$.  Next, by the Dominated Convergence Theorem, it follows that $\mathbb{E}\lrsqb{\mbs{L}_\cR(\bgamma_k,\Xs)}\approx\mathbb{E}\lrsqb{\mbs{L}_\cR^0(\bgamma_k,\Xs)}$.  Finally, by the extension of the Argmax-Continuous Mapping Theorem derived by \citet{Ferger2004}, it follows that the minimizer of $\mathbb{E}\lrsqb{\mbs{L}_\cR(\bgamma_k,\Xs)}$ will also converge a.s. to the minimizer of $\mathbb{E}\lrsqb{\mbs{L}_\cR^0(\bgamma_k,\Xs)}$, hence the order chosen by the MPM is optimal for prediction (QED).

\bigskip
The conditions under which Theorem 1 holds are quite general, as $\mathbb{E}\lrsqb{\omega|\by, \MF}$ commonly converges to a finite non-negative value by the usual consistency results \citep[e.g., see][]{Liang2008, Womack2014}, and additionally, from \citet{Tenbusch1997} we know that $\cR$ should be at most $O(n^{2/3})$, such that the condition $\cR/n\rightarrow 0$ is met in practice.  Although $\cR^\star$ as defined in \eqref{eq:Rstar} may be difficult to compute, by using Theorem 1, we obtain an approximate order $\hat{\cR}^\star$ by using the MPM.  Next we explore the performance of $\hat{\cR}^\star$ using several simulated data scenarios.

\section{Simulation Study}

In this section we test the ability of the proposed selection algorithm, choosing the order of the BPs using several simulated data scenarios. Additionally, we contrast these results to those obtained by 5-fold cross validation (CV) with functions from the {\sf cvTools} package in R.

In particular, we ran two distinct sets of simulations.  The first one takes into account all combinations of sample size ($n=100,200,500$), signal-to-noise ratio (SNR=$0.5, 1, 2$ defined later) and two true mean functions $\mu(\cdot)$. With each combination of these, we drew at random 100 datasets with one predictor and it's corresponding response.  
The second set of simulations is aimed at determining whether or not, when compared to cross-validation, the proposed strategy yields similar models in terms of accuracy and parsimony, and if our method is computationally as efficient.  For this set of simulations 100 datasets  were drawn assuming $n=10^4$, SNR$=2$ and only one true mean function.

For each simulated dataset, we first generated $x_i\sim\text{U}\lrsqb{a,b}$, and then the response $y_i \sim \text{N}(\mu(x_i),\sigma^2)$, with $\sigma$ determined by the SNR level chosen. The definition considered for the SNR, with $x\in\lrsqb{a,b}$, is 
\bean
\text{SNR}&=&\frac{1}{\sigma}\int_a^b \frac{\left|\mu(x)\right|}{b-a} dx.
\eean

The first mean function relates the outcome to the predictor through an order five polynomial.  The second, corresponds to a piecewise linear function, which is more challenging as it is not differentiable everywhere. Both our method and cross-validation were applied to every dataset to select the order of smoothness.  In every case we assumed the full model (largest possible model) to be of order $\cR=\lfloor n^{2/3}\rfloor$. The results were assessed in terms of:
\begin{description}
\item[Computational speed:] the time (in seconds) taken by each method per dataset
\item[Model complexity:] frequency counts for the selected order (out of the 100 datasets per scenario) with each method
\item[Accuracy:] the sup-norm, given by $\kappa_k=\| \tilde{\mu}_k(\cdot)-\mu(\cdot)\|_\infty$, as defined in Section 2.  This norm was chosen as it dominates all other $L_p$ norms (i.e., $\|\tilde{\mu}(\cdot)-\mu(\cdot)\|_p\leq \|\tilde{\mu}(\cdot)-\mu(\cdot)\|_\infty$ for $p\geq1$). The sup-norms considered were
\begin{itemize}
\item $\| \tilde{\mu}(\cdot)-\tilde{\mu}_\cR(\cdot)\|_\infty$, to see the performance of the largest order model considered;  
\item $\| \tilde{\mu}_{\cR^\star}(\cdot)-\mu(\cdot)\|_\infty$, where $\cR^\star$ corresponds to the order of the MPM resulting from the selection;  
\item $\| \tilde{\mu}_{\cR^{\text{cv}}}(\cdot)-\mu(\cdot)\|_\infty$, where $\cR^{\text{cv}}$ is the order of smoothness chosen by 5-fold cross-validation.
\end{itemize}

\end{description}

\subsection{Overall performance}

\begin{figure}[htbp]
\centering
\subfigure[Polynomial]{\includegraphics[scale=0.45]{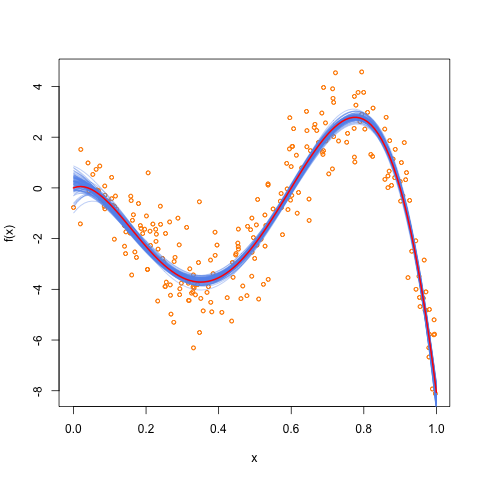}}
\subfigure[Piecewise linear]{\includegraphics[scale=0.45]{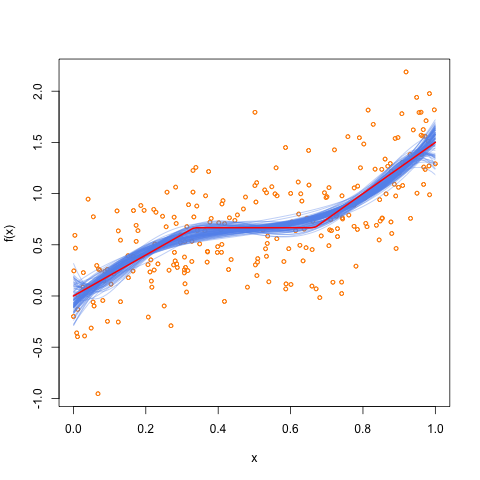}}
\caption{Functions considered for simulated data experiments. True mean function in red, simulated data in orange, and fitted curves using MPM in blue}\label{fig:TrueMeanFns}
 \end{figure}

The first mean function used is the order 5 polynomial given by $\mu(x)=5x(5x-0.2)(0.4x-1.8)(3x-1.8)(2x-1.8)$, and the second one is the piece-wise linear function $\mu(x)=x \text{I}_{\lrsqb{-3,-1}}(x)-\text{I}_{\lrsqb{-1,1}}(x)+(x-2)\text{I}_{[1,3)}(x)$ (see Figure \ref{fig:TrueMeanFns}).  With both functions, the fitted curves for the 100 datasets generated from these models matched closely the true curves, regardless of the sample sizes and levels of SNR (see Figures \ref{fitf1} and \ref{fitf4} in the appendix).  For the particular order 5 polynomial chosen, the true mean function has three stationary points, implying that it can be approximated by an order four polynomial. Interestingly,  in this case the MPM commonly chose the order to be 4 for all sample sizes and SNR values (see Figure \ref{postCVf1}). In spite of the fact that this is not the true model, this indicates that the method will identify a suitably parsimonious approximation. Similar values for the order of smoothness were often selected with CV; however, as the SNR or sample size increased, for some datasets CV chose polynomials of substantially larger degree.  For the piece-wise linear function, as the sample size or the SNR increased with either method the selected order increased, again spreading out to slightly larger values when using CV (Figure \ref{postCVf4} in appendix).

Even though CV is selecting more complex models as SNR or $n$ grow with both mean functions $\mu(\cdot)$ used in our simulations, the accuracy of these models does not improve that from the MPM's, as evidenced in Figures \ref{supnCVf1} and \ref{supnCVf4} in the Appendix.  Finally, in either case, it is also clear that simply using the full order $\cR$ model overfits the data, as it produces the largest predictive errors among the three orders of smoothness compared, namely $\cR^\star, \cR^{\text{cv}}$ and $\cR$. Lastly, the computational gains from using the  proposed approach when compared to CV are astonishing, reducing the computational time by one or two orders of magnitude (see Tables \ref{tab:timef1} and \ref{tab:timef4} in the Appendix).

\subsection{Comparison with cross-validation under ideal conditions}

\begin{figure}[htbp]
\centering
\subfigure[sup-norm]{\includegraphics[scale=0.6]{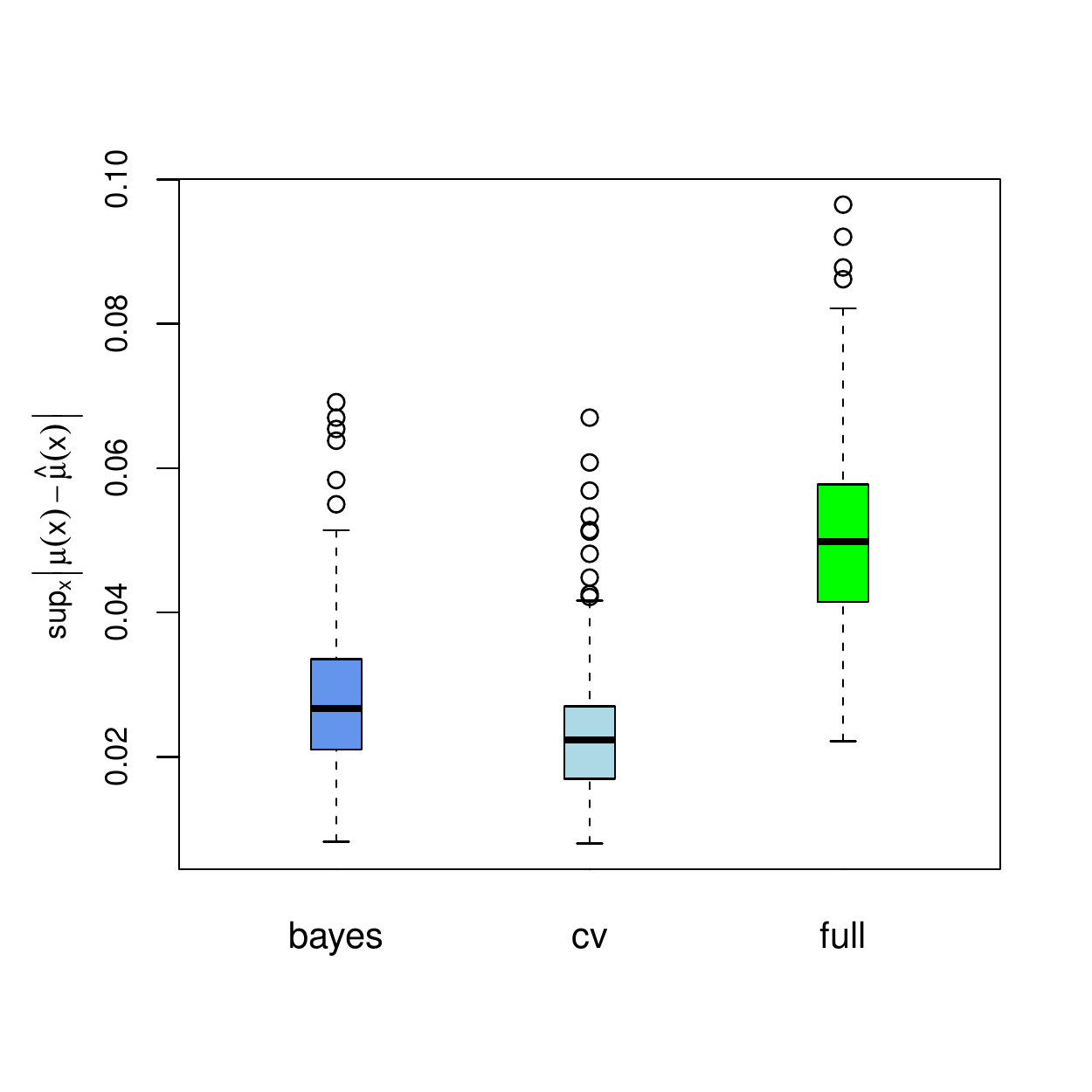}}\quad
\subfigure[computation time]{\includegraphics[scale=0.6]{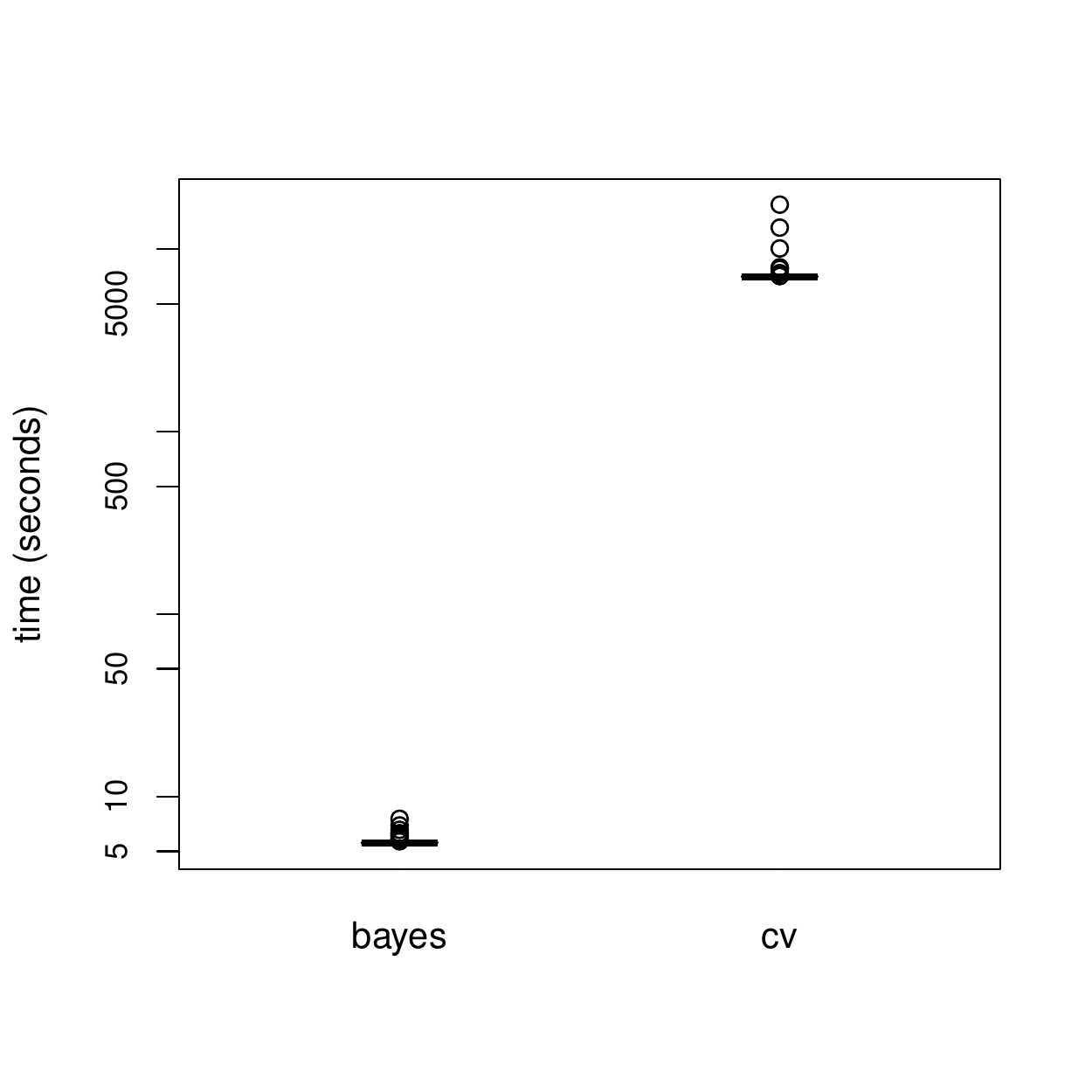}}\\
\subfigure[selected order of smoothness]{\includegraphics[scale=0.6]{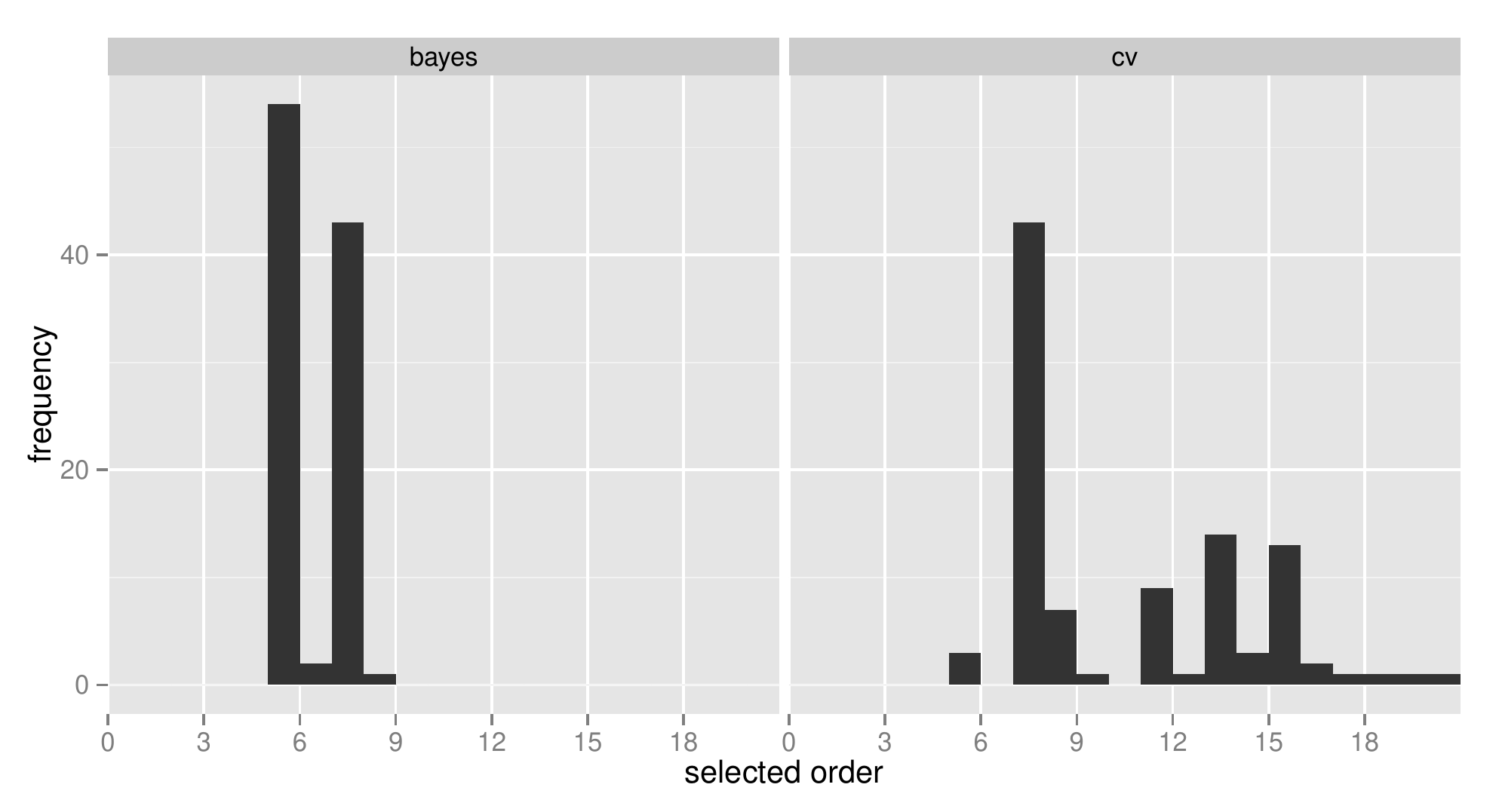}}
\caption{sup-norm error, computation time and frequencies for the selected order of the mean function $\mu(x)=x \text{I}_{\lrsqb{-3,-1}}(x)-\text{I}_{\lrsqb{-1,1}}(x)+(x-2)\text{I}_{[1,3)}(x)$ in 100 datasets, using the proposed method (bayes) and cross-validation (cv).} \label{bigsim_CVf4}
\end{figure}

Given the results from the set of simulations described in the previous section, the intent behind this set of simulation is to settle if there are in fact any significant differences between the results obtained choosing the order using the MPM and that chosen by CV.  The conditions used for this simulation were ideal in terms of having a very large sample size and considerably large signal so that good models would be easily identified.  Again, we used the piece-wise linear function $\mu(x)=x \text{I}_{\lrsqb{-3,-1}}(x)-\text{I}_{\lrsqb{-1,1}}(x)+(x-2)\text{I}_{[1,3)}(x)$. The conclusions are unequivocal:  the proposed method, when compared to CV, is comparable in predictive ability, chooses more parsimonious models, and does so in a fraction of the time taken by CV (Figure \ref{bigsim_CVf4}). In addition, the results for the full order $\cR$ were also included in our analysis providing clear indication of data overfitting, this was a recurring theme in all of the simulation experiments performed throughout the section.

\section{Data examples}

In this section, we select the order of smoothness $\cR^\star$ for the estimated mean function $\tilde{\mu}_{\cR^\star}(\cdot)$ with two real datasets using Bernstein polynomials. One of the datasets has continuous response and the other one has a binary one.  We also describe succinctly the strategy used to adapt the method for binary responses.  It is important to highlight that in this section  we provide the distribution for the order of smoothness, with which we can quantify the uncertainty associated to selected order.  This is not possible using CV; in the previous section we were able to provide frequency plots given that each scenario was analyzed on 100 datasets.

\subsection{Selecting $\cR^\star$ with continuous response}

For this example we use the child growth data from \citet{Ramsay1998}, which was also previously analyzed in \citet{McKayCurtis2011}. These data contain 83 height measurements from a 10 year old boy over a period of 312 days. In this analysis, we compare the selected order $\cR^\star$ BP model, to the mean function estimated using Bayesian isotonic regression method \citep{McKayCurtis2011} found in the {\sf bisoreg} package in R.

\begin{figure}[htbp]
\centering
\mbox{\subfigure[$\tilde{\mu}_{\cR^\star}$ and $\tilde{\mu}_{\cR}$]{\includegraphics[scale=0.35]{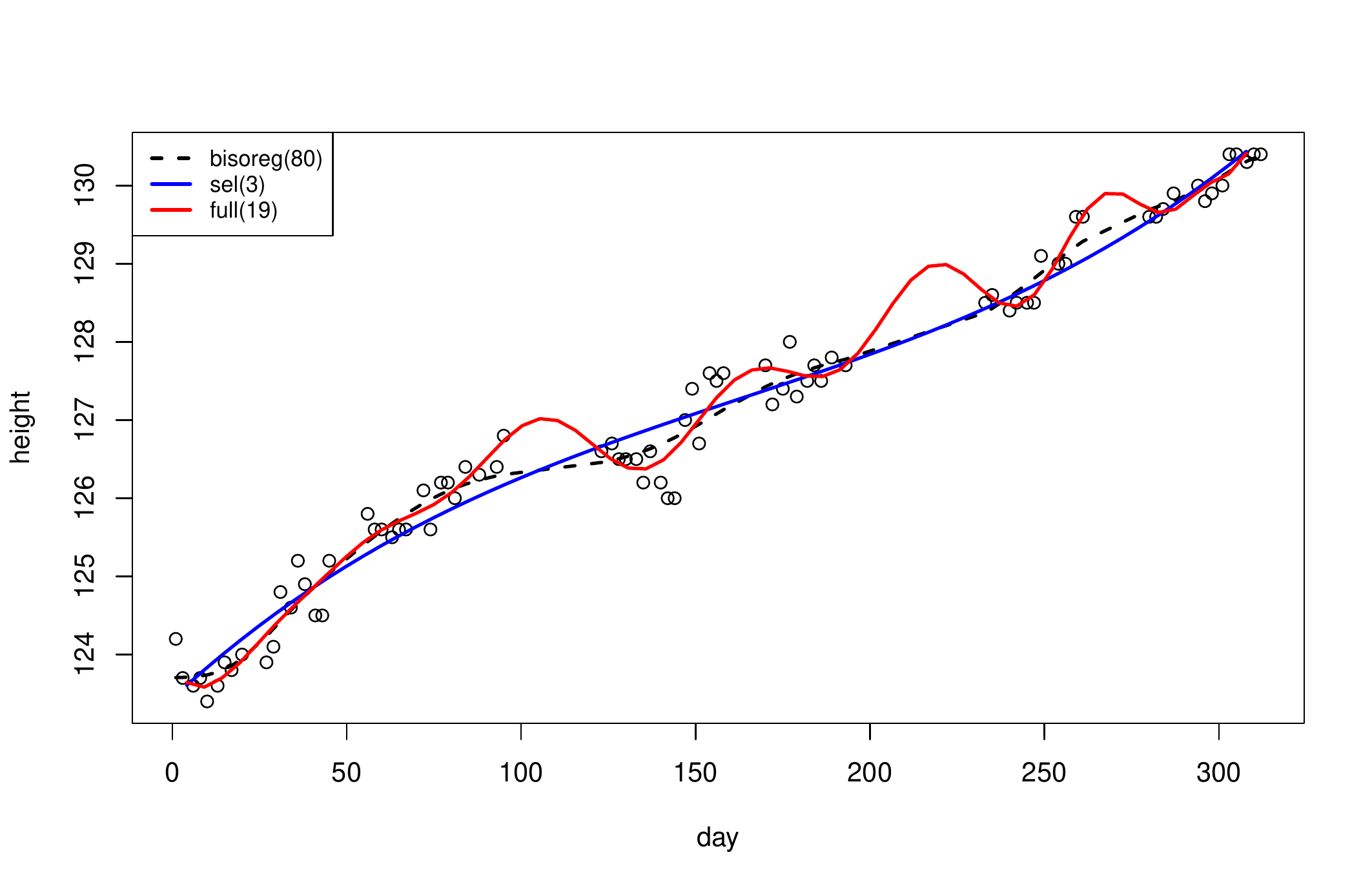}}\qquad \subfigure[$p(\cR^\star|\by$): posterior distribution $\cR^\star$]{\includegraphics[scale=0.35]{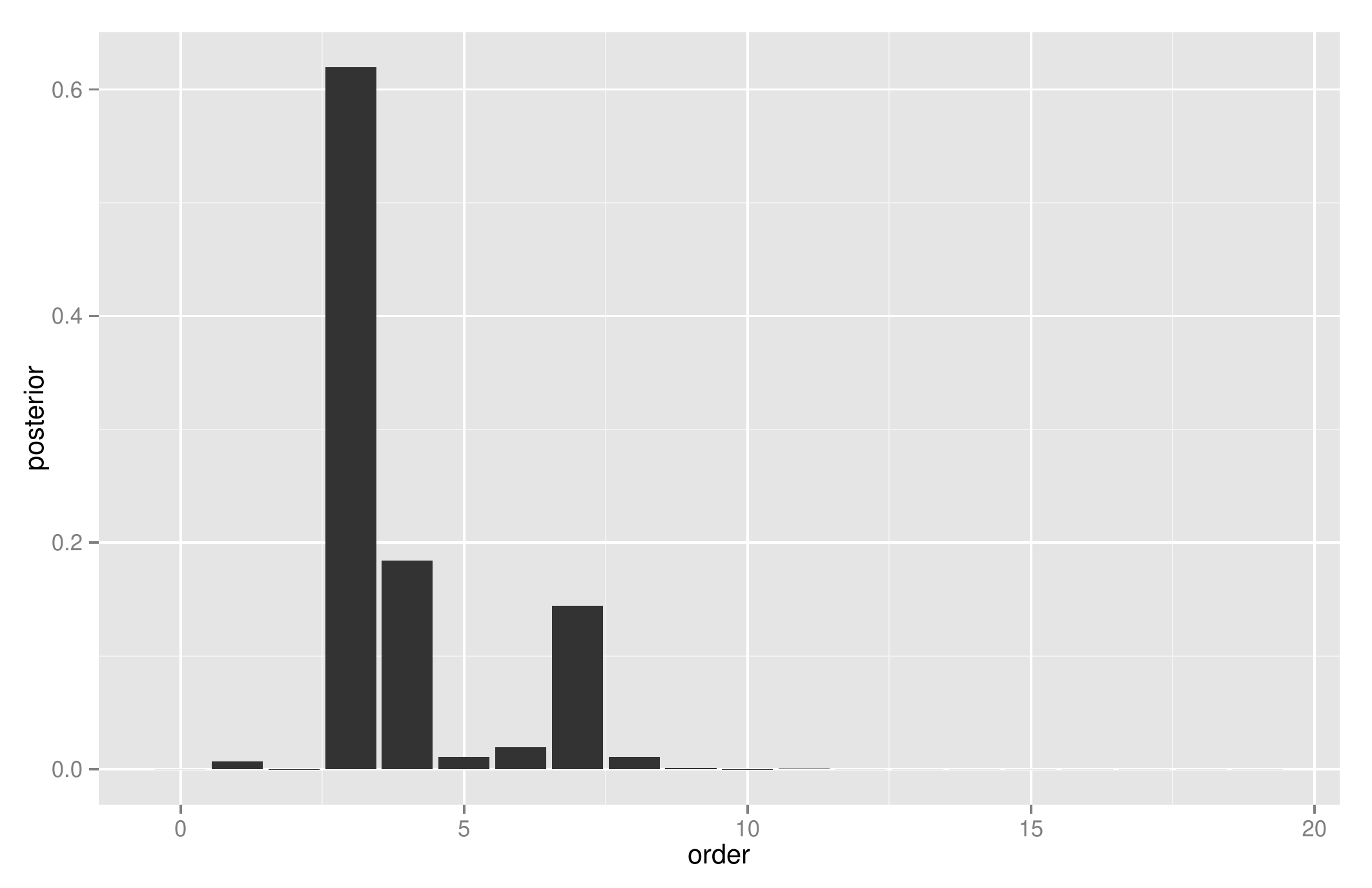}}}
\caption{Child height vs days.  Lines in blue and red correspond to the fitted means using the order $\cR^\star$ and $\cR$ BP models and the dashed line corresponds to fit using Bayesian isotonic regression with BPs (left). Posterior probability for the order of smoothness $\cR^\star$ (right).}\label{Growth}
\end{figure}

In this example there is a fair degree of certainty in the choice for the order of smoothness, as more than 60\% of the distribution concentrates on $\cR^\star=3$ when we choose $\cR=\lfloor n^{2/3}\rfloor=20$.  A reasonable assumption for this type of data is that it is non-decreasing; however, the full model fit again goes against common sense.  As such, we compare the results from our procedure to those from an order-80 Bayesian monotonic BP regression.  Remarkably, the parsimonious model selected with our method, provides a fit very close to that of {\sf bisoreg}, with $\text{sup}_{x\in\lrsqb{0,1}}\big| \tilde{\mu}_{\cR^\star}(x)-\tilde{\mu}_{\text{bisoreg}}(x)\big|=0.2541$, which is outstanding considering that the observed range for the responses is between 123 and 131.

\subsection{Selecting $\cR^\star$ with binary response}

In order to deal with binary responses we follow the strategy proposed in \citet{Leon-Novelo2012} with intrinsic priors, which uses data augmentation to introduce latent normal random variables that make the problem tractable.  The approach builds upon the sampling algorithm of \citet{ACh_93} for probit regression models.  This selection problem is solved on the latent scale, since each model $\bgamma\in\cM$ can be viewed as a normal regression model where only the sign of the response is observed. Explicitly, associated to the response vector $\by=(y_1,\ldots,y_n)$, where $y_i|\bgamma_k\sim\text{Bernoulli}\lrp{\Phi\lrp{\bz_k(x_i)^\prime \blambda _k}}$, we may consider a sample of latent random variables $\bv=(v_{1},\dots ,v_{n})$, where  $v_{i}|\bgamma_k\sim N(v_i|\bz_k(x_i)^\prime \blambda _k,1)$.  Each of the observed responses and the latent variables are connected through the equality $y_{i}=I(v_{i}>0)$. 

In the binary response case the goal is to derive the marginal probabilities for the observed responses from the marginal densities calculated for the latent variables.  The marginal density of $\by$ under model $\bgamma_k$, are given by 
\bea
m(\by|\bgamma_k)&=&\int_{A_1\times\ldots\times A_n} m(\bv|\bgamma_k)d\bv\nonumber\\
&=&\int_{A_1\times\ldots\times A_n} \lrp{\int_{-\infty}^{\infty}m(\bv|\bgamma_k,\lambda_0) d \lambda_0} d\bv \nonumber\\
&=& \int_{-\infty}^{\infty} \underbrace{\lrp{\int_{\mbf{A}}m(\bv|\bgamma_k,\lambda_0) d\by}}_{g_k\lrp{\mbf{A}| \bgamma_k,\lambda_0}} d \lambda_0  \label{eq:BernIP}
\eea
where $\mbs{A}=A_1\times\ldots\times A_n$ and $A_i=(0,\infty)$ if $y_i=1$ and $A_i=(-\infty,0]$ otherwise. From equation \refeq{eq:BernIP} it follows that the Bayes factor for each model $\bgamma\in\cM$ in the binary case is given by
\bea
BF_{\bgamma_k,\MB}(\by)&=&\frac{\int_{-\infty}^{\infty} g_k\lrp{\mbf{A}| \bgamma_k,\lambda_0} d \lambda_0}{\int_{-\infty}^{\infty} g_{0}\lrp{\mbf{A}| \bgamma_0,\lambda_0} d \lambda_0},
\eea
with
\bean
g_k\lrp{\mbf{A}|\bgamma_k, \lambda_0}&=&c_0 \int_{\mbf{A}} \mathcal{N}_{n}\lrp{\bv | \lambda_0 \1_n, \Sigma_k} d\by,\\
g_{0}\lrp{\mbf{A}|\MB, \lambda_0}&=&c_0 \int_{\mbf{A}} \mathcal{N}_{n}\lrp{\bv | \lambda_0 \1_n, \mbf{I}_n } d\by,
\eean
where $\Sigma_k=\mbf{I}_n+\frac{2n}{k+1}\bZ_k\lrp{\bZ_k^\prime\bZ_k}^{-1}\bZ_k^\prime$. The model posterior probabilities are then obtained plugging in the Bayes factor as in Equation \eqref{eq:postBF}.

To illustrate the effectiveness of the method using the strategy of \citet{Leon-Novelo2012}, we consider data from the experiment described in \citet{Martinez2012}. This dataset relates the incidence of metritis in lactating cows, an early postpartum disease, to the levels of calcium in the blood.  In this experiment, the authors were interested in demonstrating that subclinical hypocalcemia in dairy cows (defined as Ca<8.59 mg/dL) increases the likelihood of disease, and in particular of metritis, during the early postpartum.  The dataset includes measurements taken the third day after parturition from 73 multiparous cows in a dairy farm.

\begin{figure}[htbp]
\centering
\mbox{\subfigure[$\tilde{\mu}_{\cR^\star}$ and $\tilde{\mu}_{\cR}$]{\includegraphics[scale=0.35]{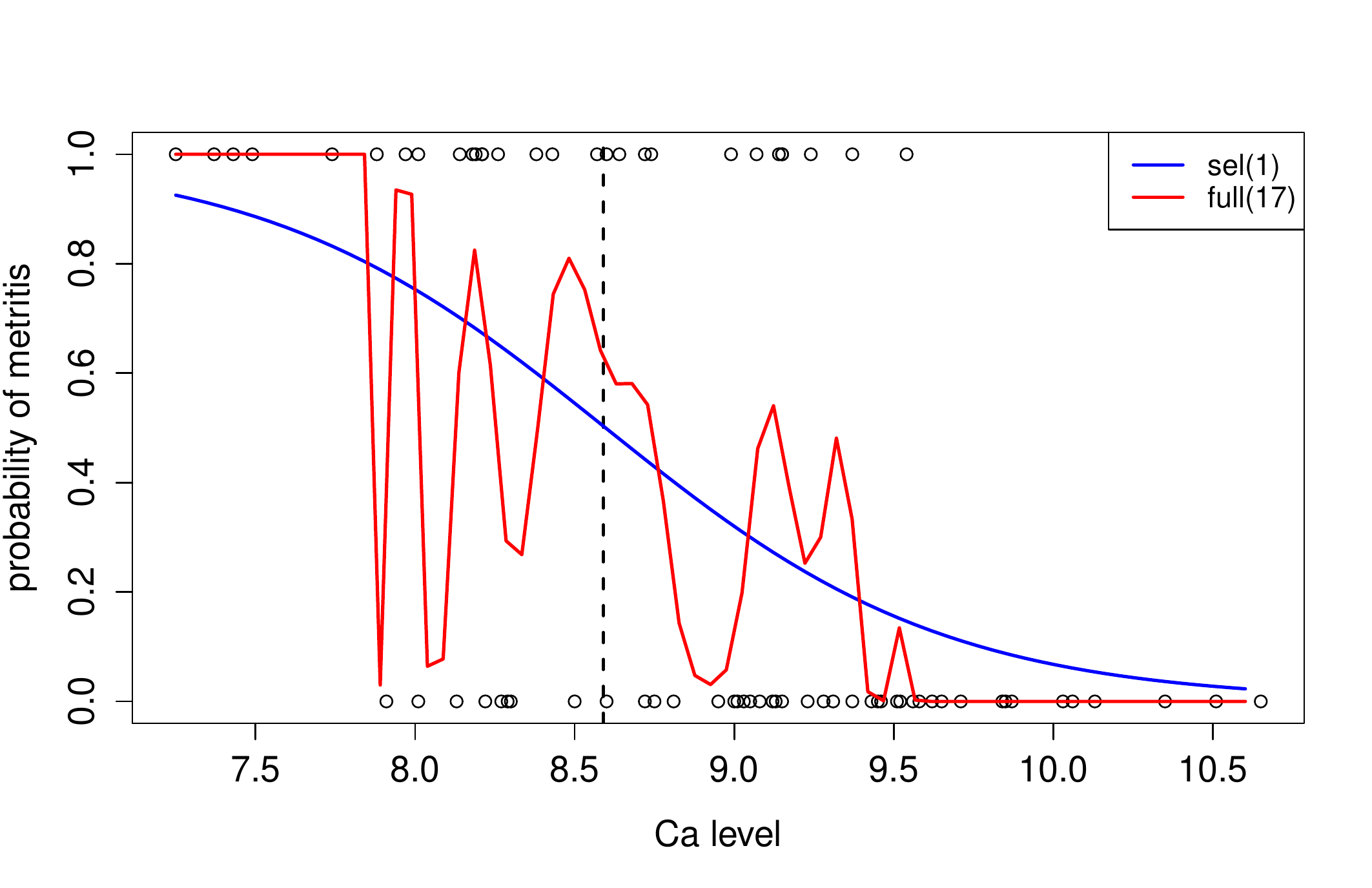}}\qquad \subfigure[$p(\cR^\star|\by$): posterior distribution $\cR^\star$]{\includegraphics[scale=0.35]{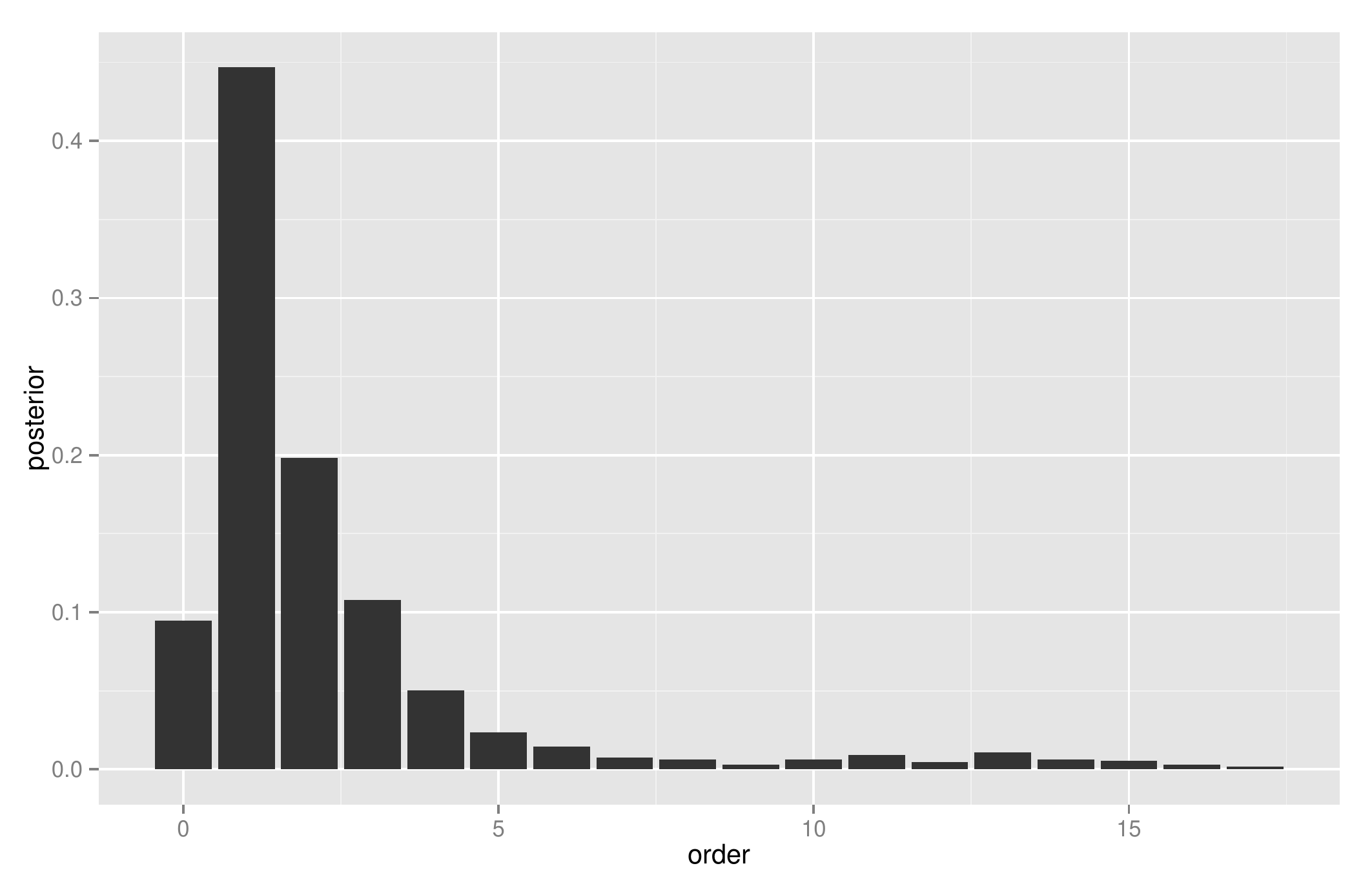}}}
\caption{Probability of metritis vs level of Calcium in blood:  Fitted means using the selected order $\cR^\star$ model and the full order $\cR$ model (left). Posterior probability for the order of smoothness $\cR^\star$ (right).}\label{CaMetritis}
\end{figure}

Figure \ref{CaMetritis} clearly demonstrates the inadequacy of using the full order $\cR$ Bernstein polynomial, which overfits the data, whereas the order $\cR^\star$ BP model (in this case order 1 in probit scale) is consistent with the intuition that the relationship between calcium and metritis is negative monotonic.  Interestingly, the point of inflection for the fitted curve with the selected order of smoothness occurs near the cutoff for subclinical  hypocalcemia (dashed line in Figure \ref{CaMetritis}).  In spite of these results, the heavy tails of the posterior distribution for $\cR^\star$ indicate that there is some uncertainty surrounding this choice.

\section{Concluding Remarks}

Using tools from the objective Bayesian variable selection literature and exploiting the fact that there is a linear map between Bernstein and orthogonal polynomial basis, we devise an efficient procedure to select optimally and automatically the order of smoothness used in the estimation of a nonparametric regression function with Bernstein Polynomials.   The functions used throughout were built into the R package {\sf AutoBPFit}. The proposed strategy was shown to be asymptotically optimal for prediction if the order of smoothness is chosen based on the MPM.  In simulation experiments, our method, when compared to CV proved to be one or two orders of magnitude faster and selected consistently more parsimonious models, remarkably, without any significant loss in predictive accuracy.   The method was extended to binary responses, and it can potentially be used with other exponential family responses where a latent normal representation is available.   Throughout this article we treated the single predictor case, making use of the immediate inheritance assumption in the construction of the model priors.  However, the method can be seamlessly extended to the multiple predictor case by enforcing the principle of {\sl conditional independence} as defined in \citet{Chipman1996}.  
In the case where several predictors are considered and selection must not only be done on the order of smoothness, but also on which predictors actually enter the model, it is commonly not possible to enumerate the entire model space, and instead a stochastic search algorithm to explore it can be considered \citep[e.g., see ][]{Taylor-Rodriguez2015}.  

\end{spacing}

\pagebreak
\bibliography{BPLit2}
 \bibliographystyle{apalike}

\pagebreak 
 \appendix
 
 \section{Figures and Tables simulations Section 4.1}
 
 \begin{figure}[htbp]
\centering
\subfigure[$n=100$]{\includegraphics[scale=0.5]{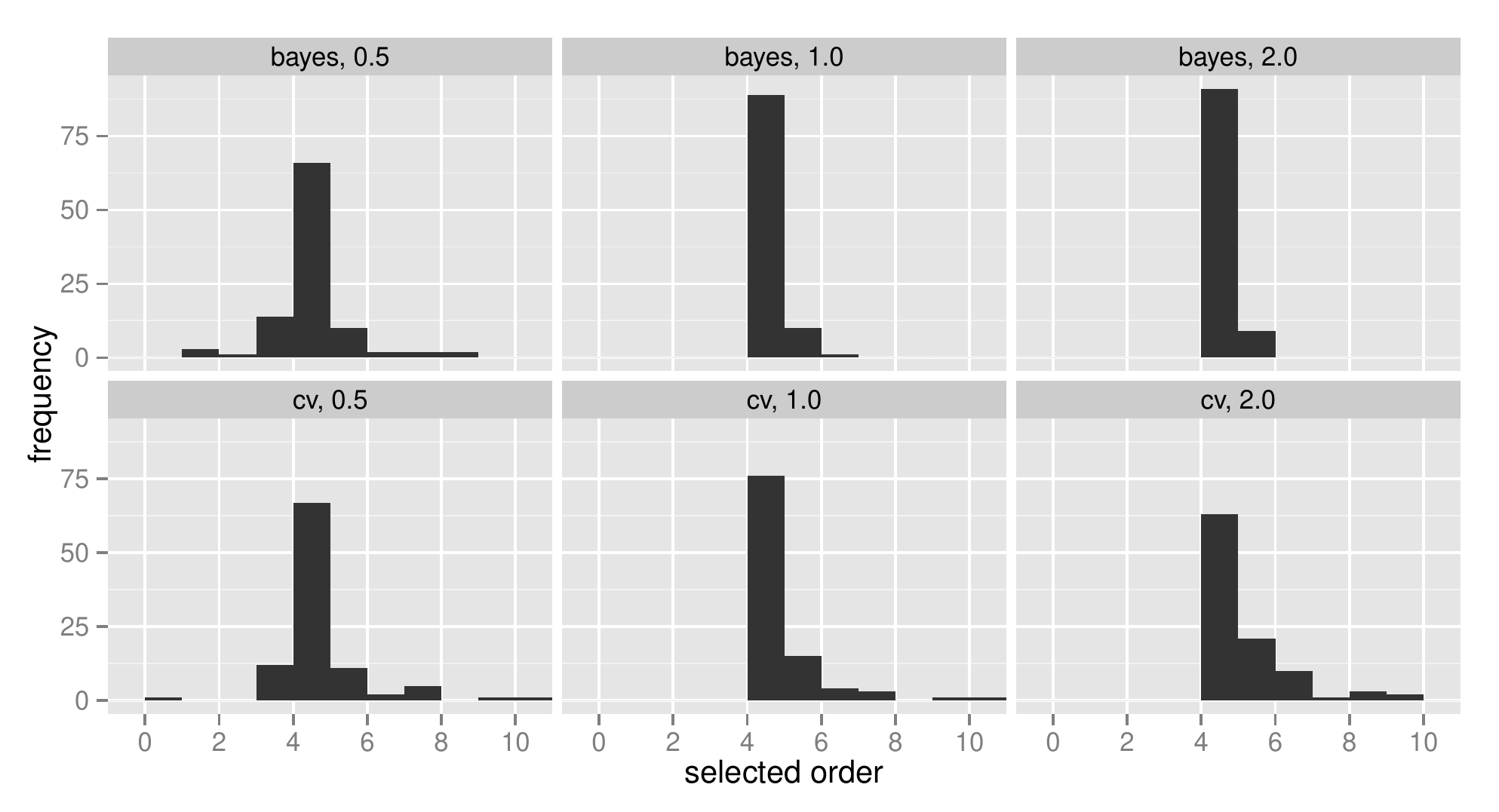}}\\
\subfigure[$n=200$]{\includegraphics[scale=0.5]{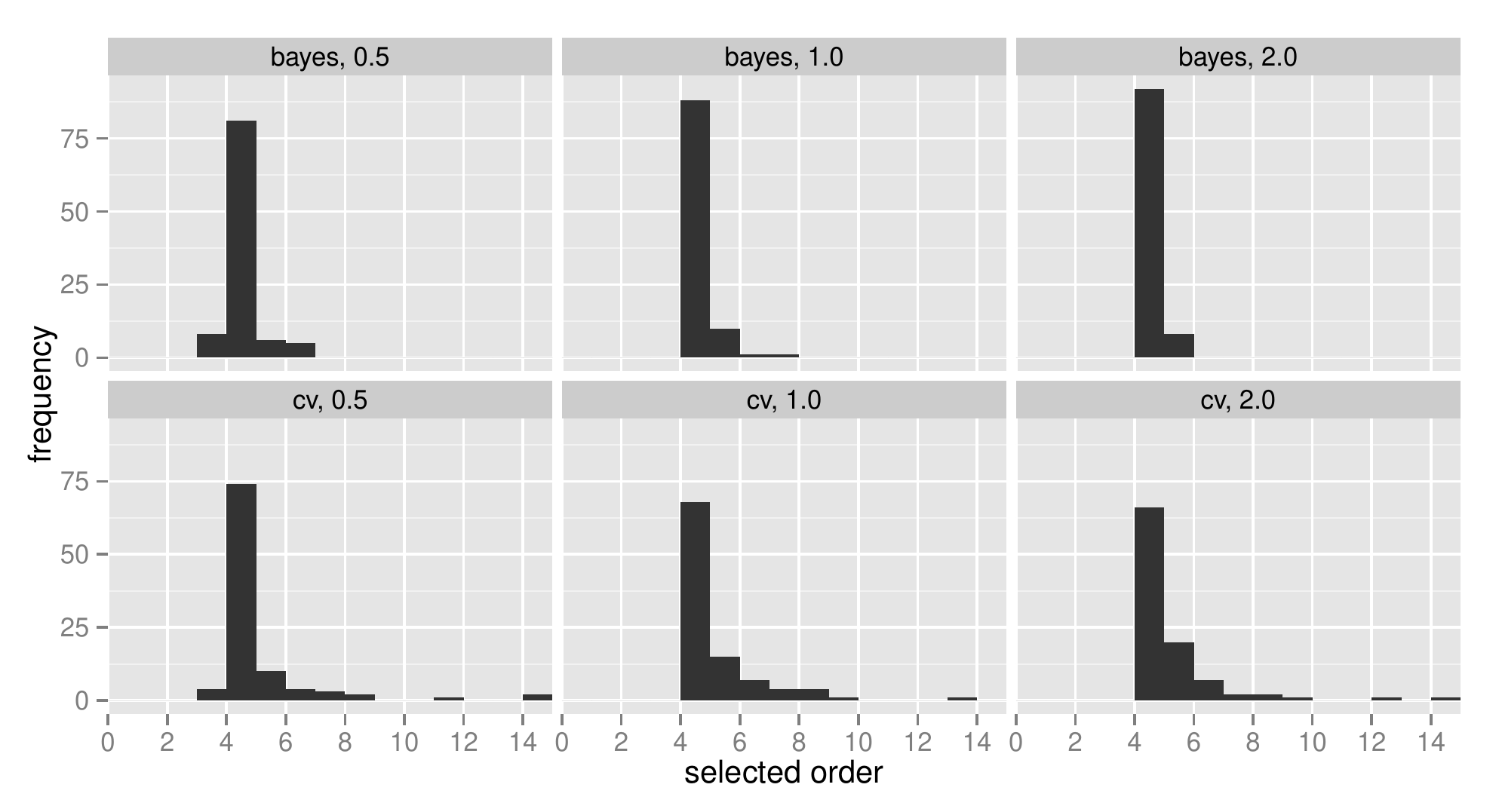}}\\
\subfigure[$n=500$]{\includegraphics[scale=0.5]{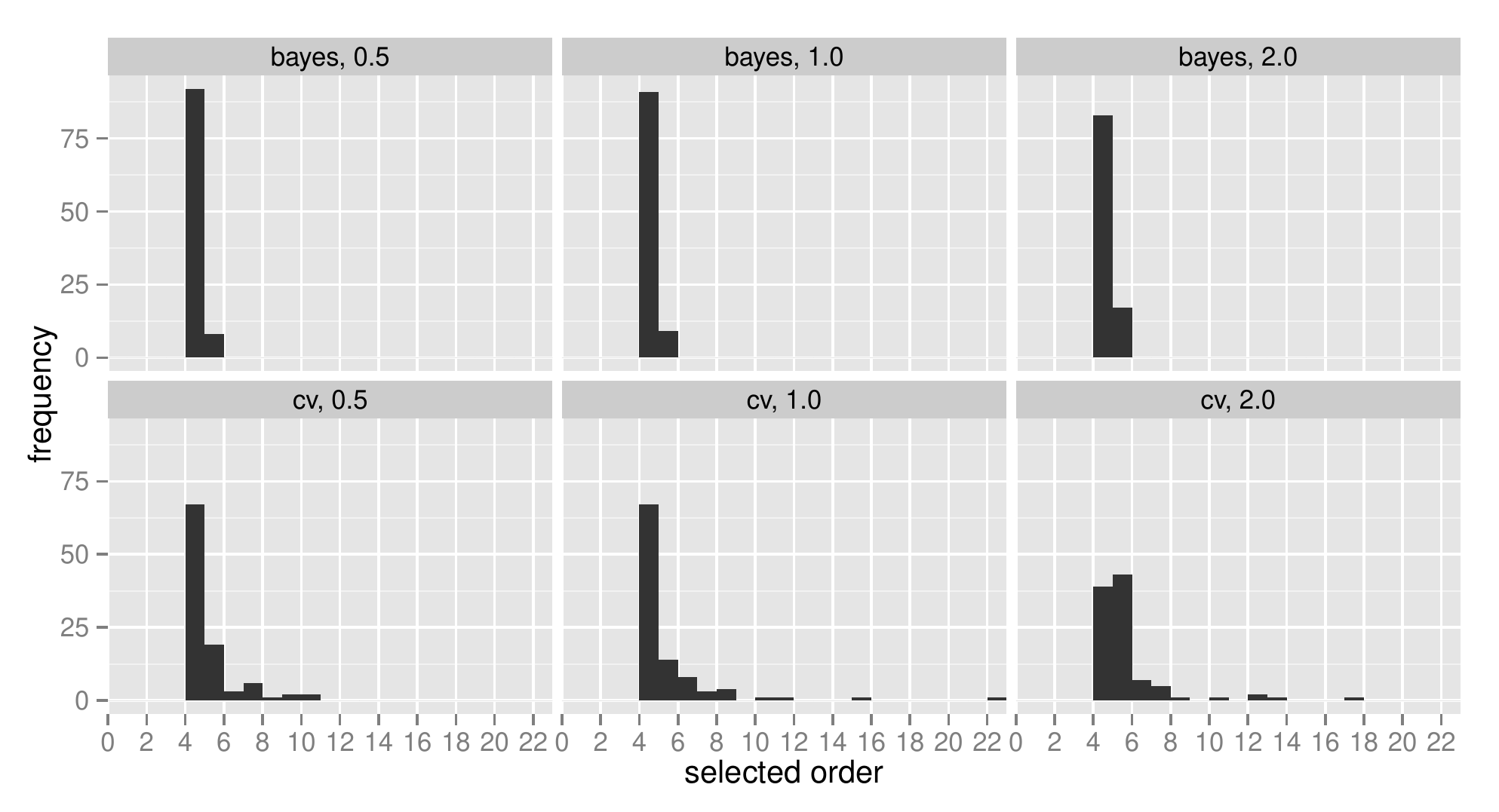}}
\caption{Frequencies for the order of smoothness selected in 100  simulated datasets with $SNR=0.5, 1, 2$ for the proposed method (bayes) and cross-validation (cv) with mean function $\mu(x)=5x(5x-0.2)(0.4x-1.8)(3x-1.8)(2x-1.8)$.} \label{postCVf1}
\end{figure}

\begin{figure}[htbp]
\centering
\mbox{\subfigure[$n=100$]{\includegraphics[scale=0.45]{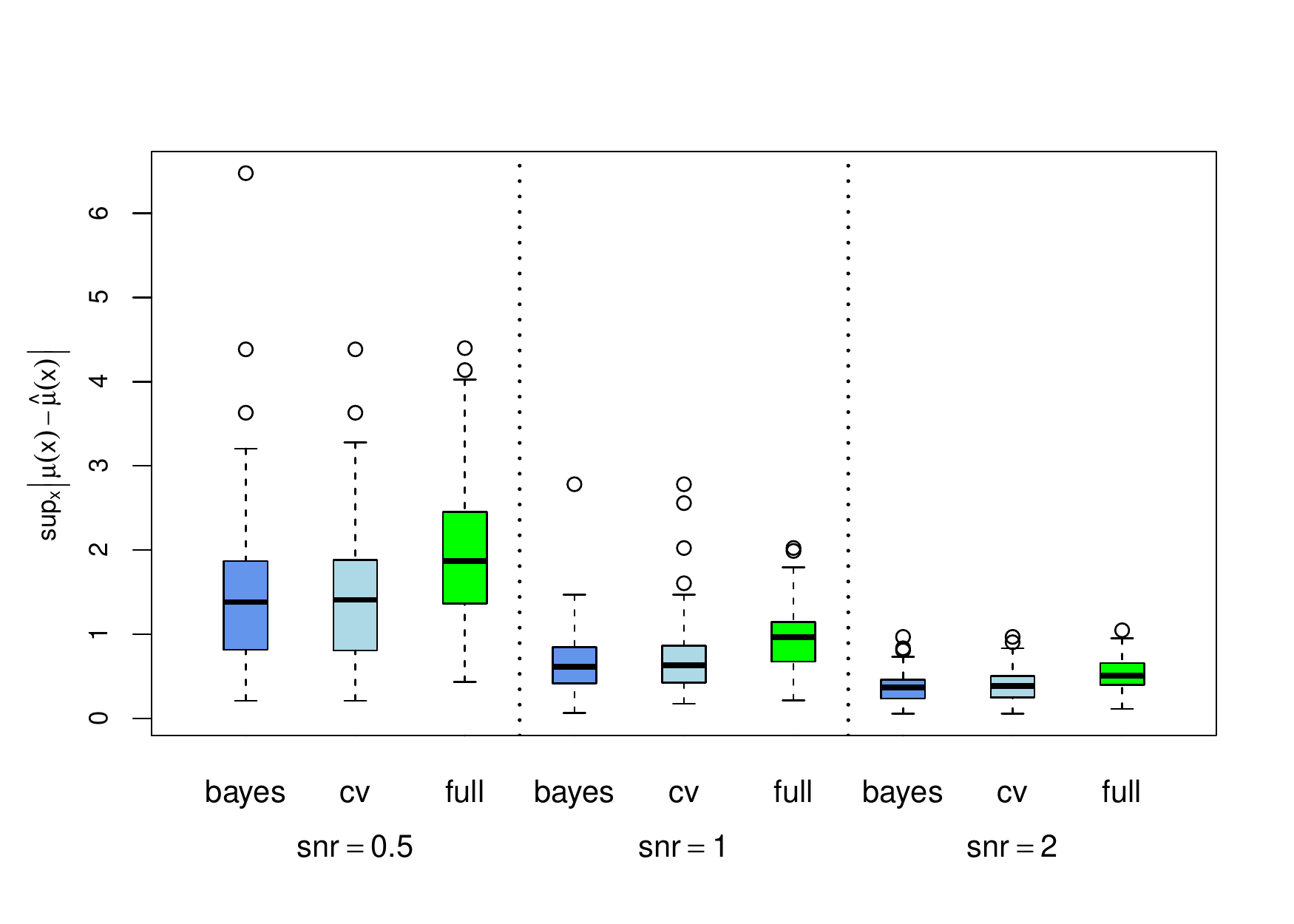}}\quad
\subfigure[$n=200$]{\includegraphics[scale=0.45]{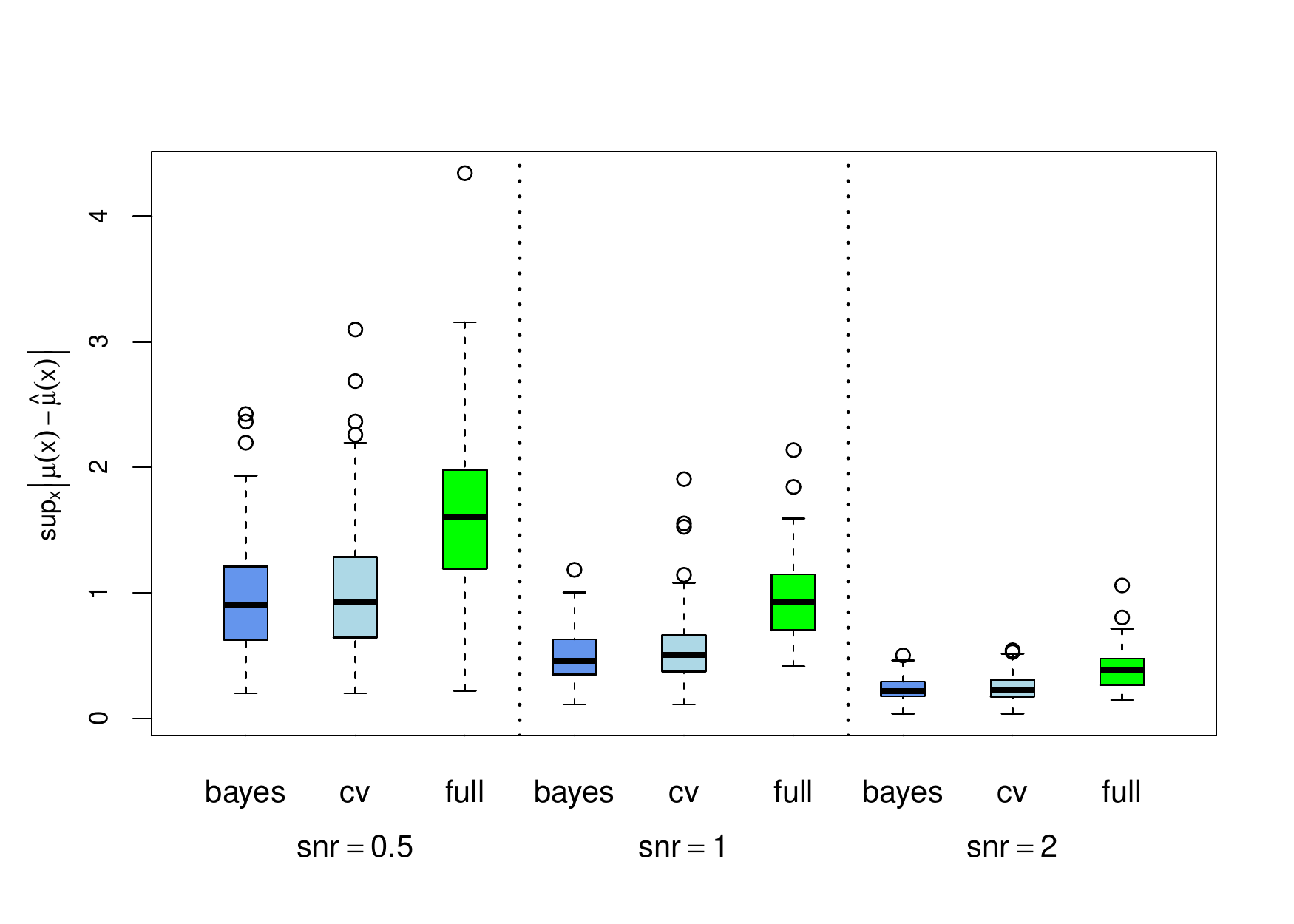}}}\\
\subfigure[$n=500$]{\includegraphics[scale=0.45]{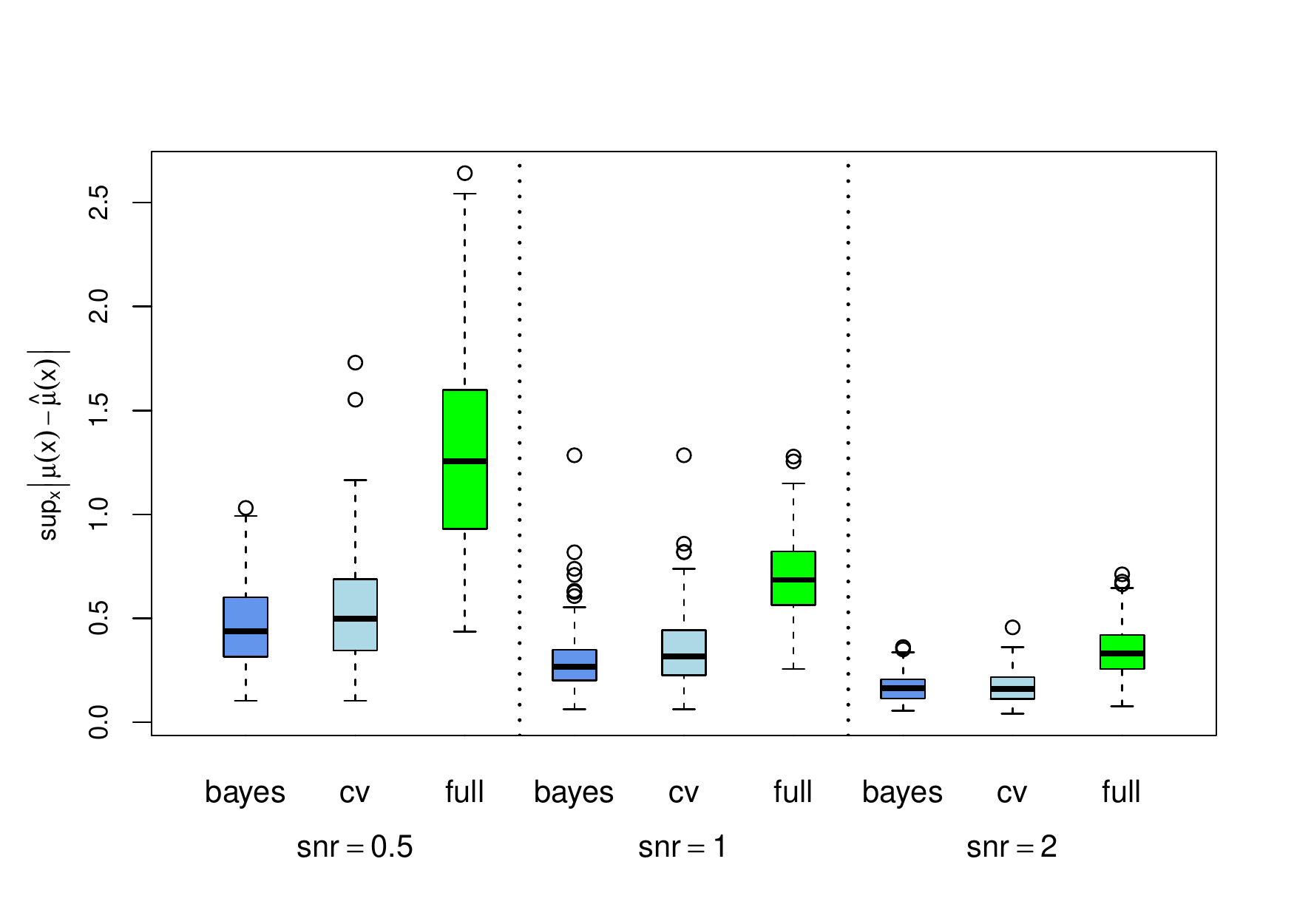}}
\caption{sup-norm for the difference between the true mean $\mu(x)=5x(5x-0.2)(0.4x-1.8)(3x-1.8)(2x-1.8)$ and the fitted values of 1) the selected order $\cR^\star$ polynomial with the proposed method (bayes), 2) the selected polynomial using cross-validation (cv), and 3) the full order $\cR$ model (full).} \label{supnCVf1}
\end{figure}

\begin{table}[htbp]
\centering
\begin{tabular}{lc | rrr | rrr | rrr}
  \toprule
 \multirow{2}{*}{method} &  \multirow{2}{*}{snr} &  \multicolumn{3}{c}{$n=100$}&  \multicolumn{3}{|c|}{$n=200$} &  \multicolumn{3}{|c}{$n=500$}\\
 &  & 2.5\% & 50\% & 97.5\% & 2.5\% & 50\% & 97.5\% & 2.5\% & 50\% & 97.5\% \\ 
  \midrule
bayes & \multirow{2}{*}{0.5} & 0.07 & 0.08 & 0.09 & 0.08 & 0.10 & 0.11 & 0.10 & 0.12 & 0.19 \\ 
  cv & &  2.03 & 2.11 & 2.26 & 4.48 & 5.17 & 6.84 & 7.12 & 8.17 & 13.40 \\ 
    \midrule
  bayes &  \multirow{2}{*}{1}  &0.11 & 0.12 & 0.15 & 0.11 & 0.13 & 0.16 & 0.12 & 0.14 & 0.17 \\ 
  cv &  & 2.01 & 2.18 & 2.97 & 3.32 & 3.63 & 5.12 & 7.39 & 9.09 & 20.77 \\ 
    \midrule
  bayes &  \multirow{2}{*}{2}  & 0.08 & 0.12 & 0.15 & 0.08 & 0.09 & 0.11 & 0.06 & 0.08 & 0.09 \\ 
  cv &  & 1.77 & 2.03 & 3.07 & 4.52 & 5.18 & 6.66 & 5.38 & 5.95 & 8.03 \\ 
   \bottomrule
\end{tabular}
\caption{Quantiles $2.5\%,50\%,97.5\%$ for the computation time per dataset (in seconds). Comparison between BP order selection with proposed method (bayes) and cross-validation (cv) with mean function $\mu(x)=5x(5x-0.2)(0.4x-1.8)(3x-1.8)(2x-1.8)$.} 
\label{tab:timef1}
\end{table}

\begin{figure}[htbp]
\centering
\subfigure[$n=100$]{\includegraphics[scale=0.5]{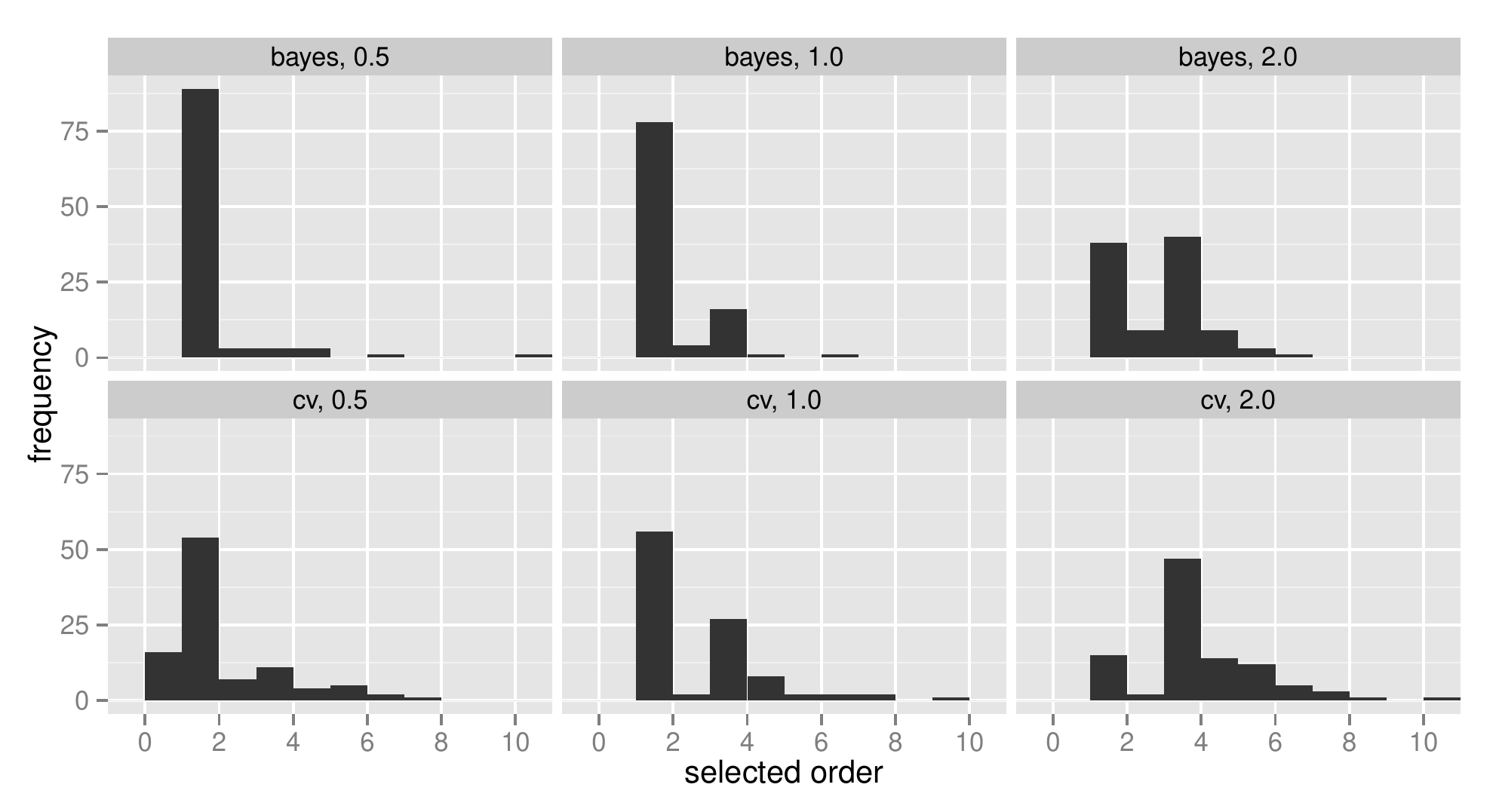}}\\
\subfigure[$n=200$]{\includegraphics[scale=0.5]{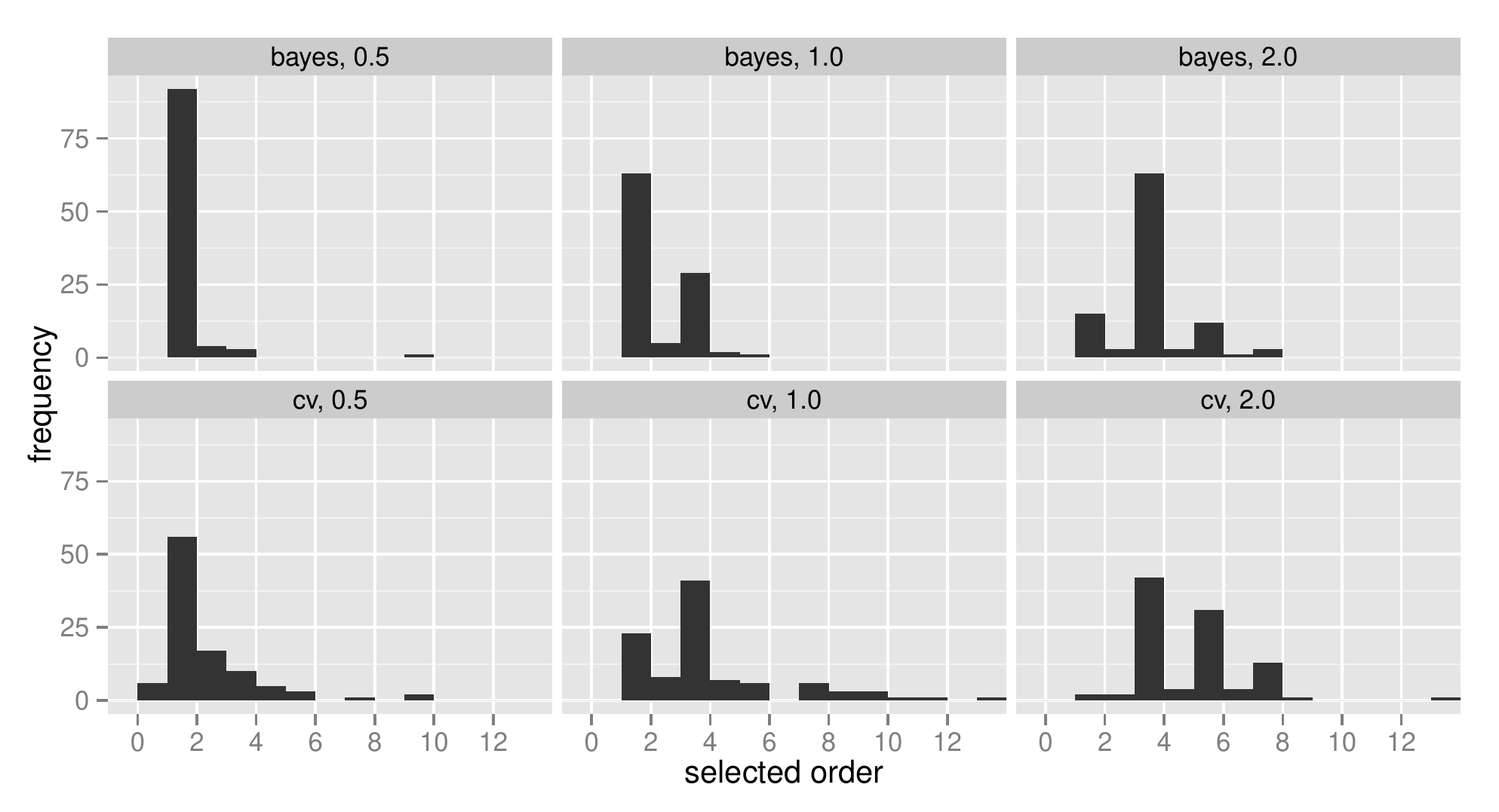}}\\
\subfigure[$n=500$]{\includegraphics[scale=0.5]{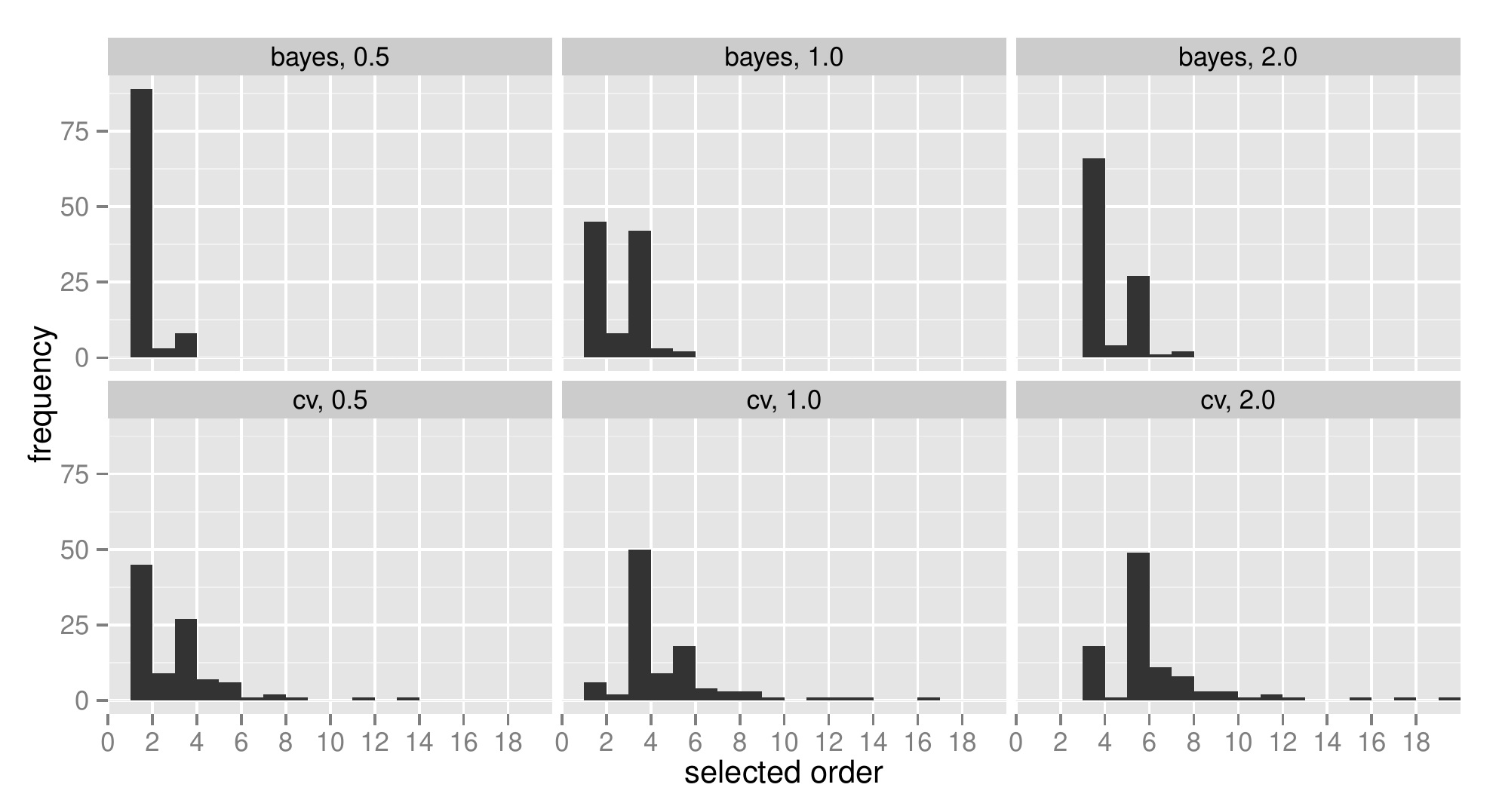}}
\caption{Frequencies for the order of smoothness selected in 100 simulated datasets with $SNR=0.5, 1, 2$ for the proposed method (bayes) and cross-validation (cv) with mean function $\mu(x)=x \text{I}_{\lrb{x\in\lrsqb{-3,-1}}}-\text{I}_{\lrb{x\in\lrb{-1,1}}}+(x-2)\text{I}_{\lrb{x\in[1,3)}}$.} \label{postCVf4}
\end{figure}

\begin{figure}[htbp]
\centering
\mbox{\subfigure[$n=100$]{\includegraphics[scale=0.45]{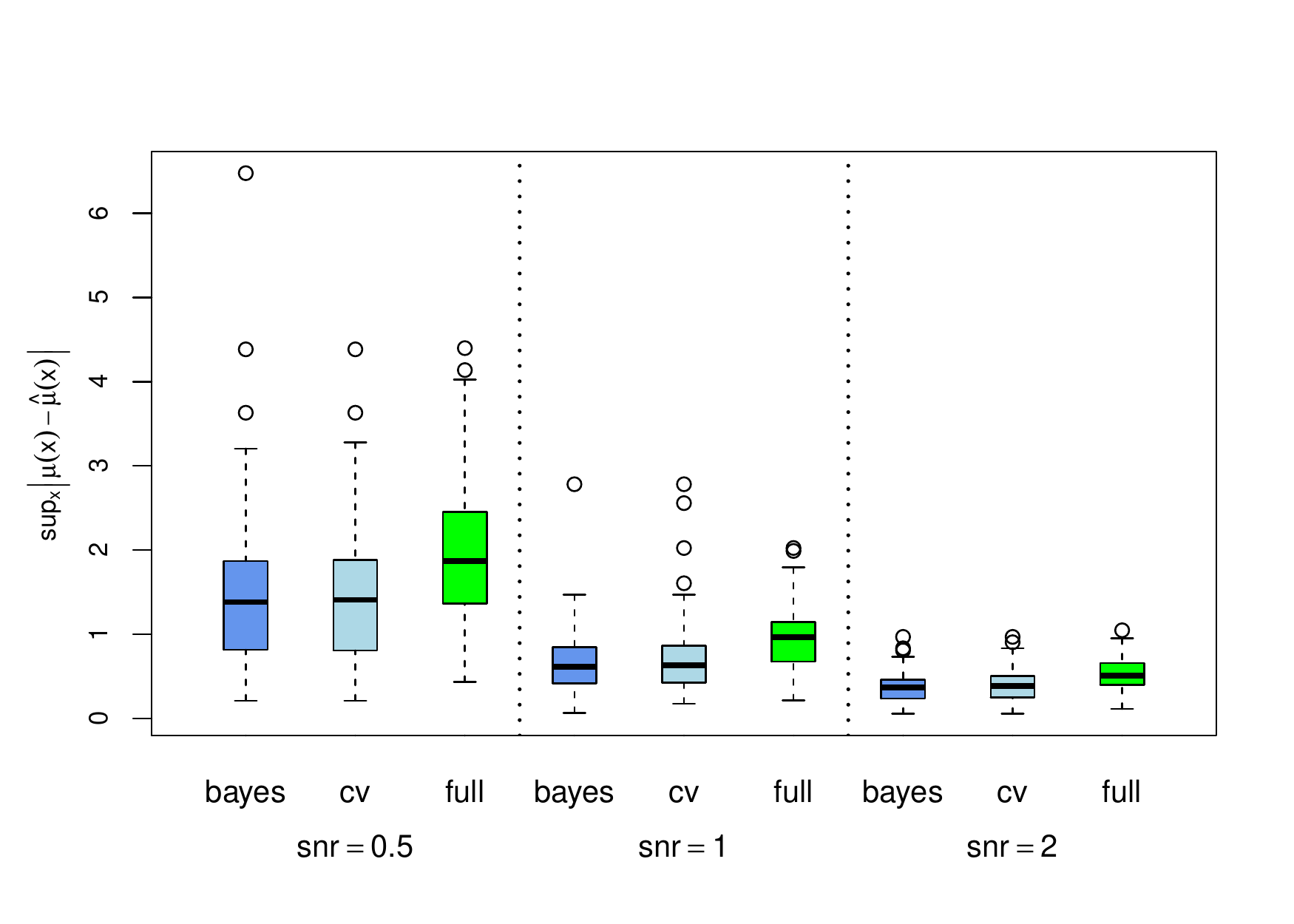}}\quad
\subfigure[$n=200$]{\includegraphics[scale=0.45]{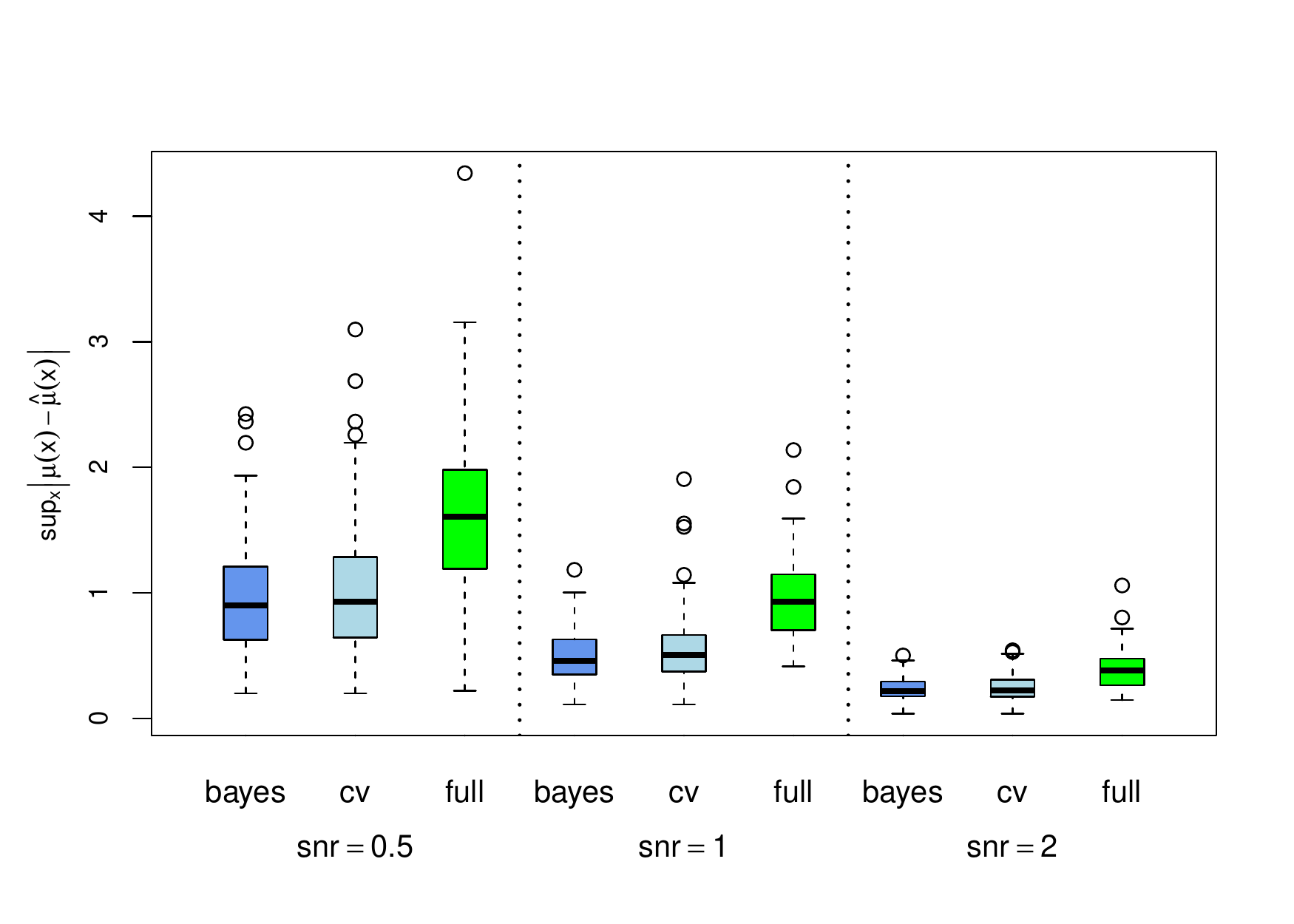}}}\\
\subfigure[$n=500$]{\includegraphics[scale=0.45]{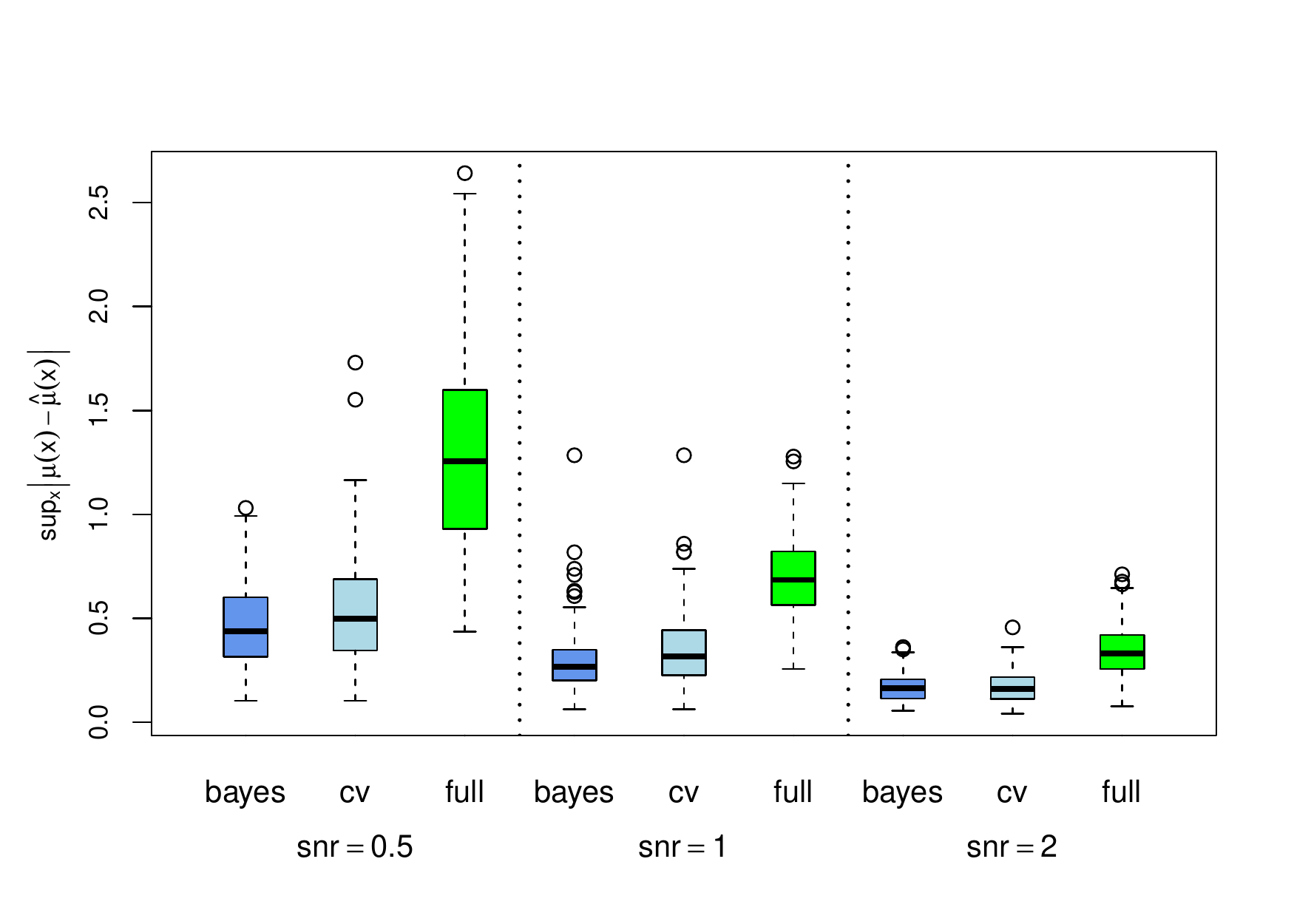}}
\caption{sup-norm for the difference between the true mean $\mu(x)=x \text{I}_{\lrb{x\in\lrsqb{-3,-1}}}-\text{I}_{\lrb{x\in\lrb{-1,1}}}+(x-2)\text{I}_{\lrb{x\in[1,3)}}$ and the fitted values of 1) the selected order $\cR^\star$ polynomial with the proposed method (bayes), 2) the selected polynomial using cross-validation (cv), and 3) the full order $\cR$ model (full).} \label{supnCVf4}
\end{figure}

\begin{table}[htbp]
\centering
\begin{tabular}{lc | rrr | rrr | rrr}
  \toprule
 \multirow{2}{*}{method} &  \multirow{2}{*}{snr} &  \multicolumn{3}{c}{$n=100$}&  \multicolumn{3}{|c|}{$n=200$} &  \multicolumn{3}{|c}{$n=500$}\\
 &  & 2.5\% & 50\% & 97.5\% & 2.5\% & 50\% & 97.5\% & 2.5\% & 50\% & 97.5\% \\ 
  \midrule
bayes &  \multirow{2}{*}{0.5}  & 0.07 & 0.08 & 0.09 & 0.12 & 0.14 & 0.16 & 0.12 & 0.14 & 0.17 \\ 
  cv &  & 2.03 & 2.10 & 2.28 & 4.05 & 4.99 & 6.87 & 7.79 & 8.99 & 20.83 \\ 
  bayes &  \multirow{2}{*}{1}  & 0.11 & 0.12 & 0.15 & 0.08 & 0.10 & 0.11 & 0.06 & 0.08 & 0.09 \\ 
  cv &  & 2.03 & 2.17 & 2.98 & 4.48 & 5.19 & 6.76 & 5.40 & 5.97 & 7.90 \\ 
  bayes &  \multirow{2}{*}{2}  & 0.06 & 0.08 & 0.09 & 0.11 & 0.12 & 0.16 & 0.07 & 0.07 & 0.08 \\
  cv &  & 2.03 & 2.12 & 2.25 & 3.33 & 3.67 & 5.24 & 5.21 & 5.38 & 6.64 \\ 
   \bottomrule
\end{tabular}
\caption{Quantiles $2.5\%,50\%,97.5\%$ for the computation time per dataset (in seconds). Comparison between BP order selection with proposed method (bayes) and cross-validation (cv) with mean function $\mu(x)=x \text{I}_{\lrb{x\in\lrsqb{-3,-1}}}-\text{I}_{\lrb{x\in\lrb{-1,1}}}+(x-2)\text{I}_{\lrb{x\in[1,3)}}$.} 
\label{tab:timef4}
\end{table}

 \begin{figure}[htbp]
\centering
\mbox{\subfigure[SNR=0.5, $n$=100]{\includegraphics[scale=0.22]{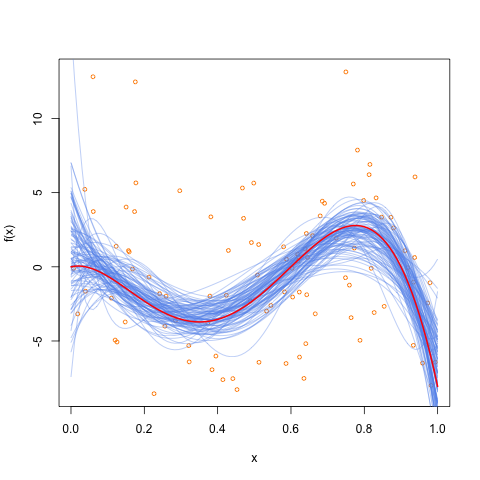}}\quad
\subfigure[SNR=0.5, $n$=200]{\includegraphics[scale=0.22]{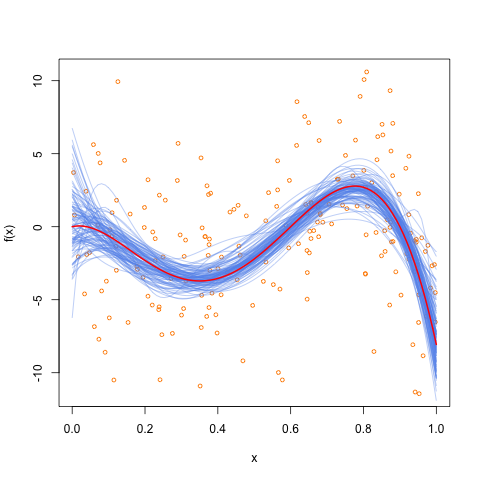}}\quad
\subfigure[SNR=0.5, $n$=500]{\includegraphics[scale=0.22]{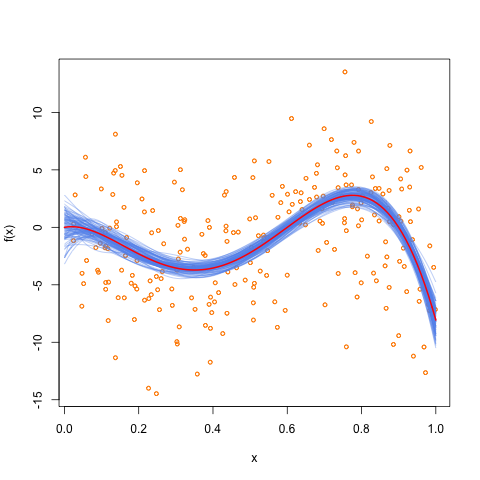}}}\\

\mbox{\subfigure[SNR=1, $n$=100]{\includegraphics[scale=0.22]{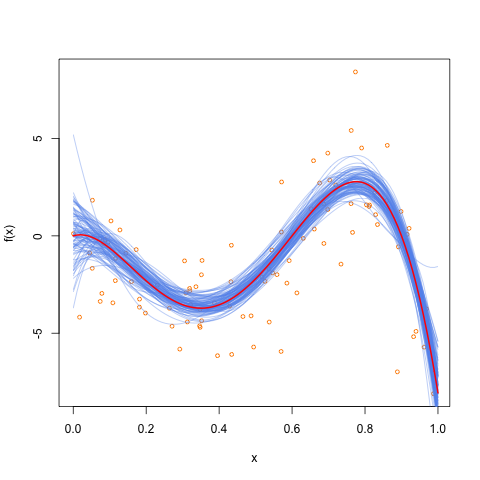}}\quad
\subfigure[SNR=1, $n$=200]{\includegraphics[scale=0.22]{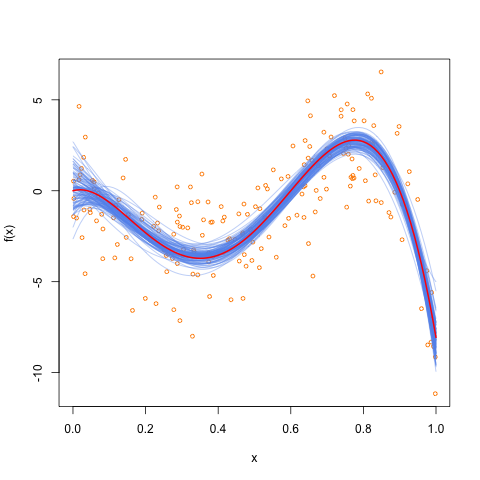}}\quad
\subfigure[SNR=1, $n$=500]{\includegraphics[scale=0.22]{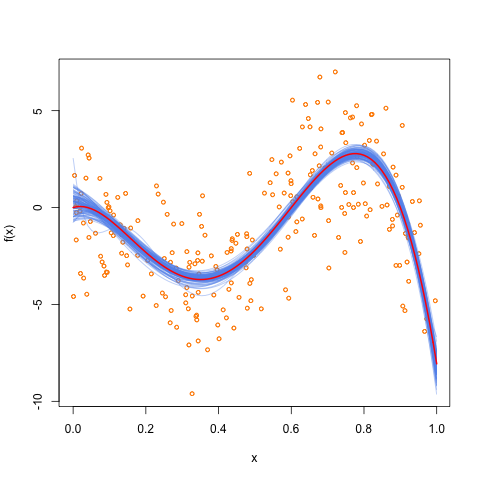}}}\\

\mbox{\subfigure[SNR=2, $n$=100]{\includegraphics[scale=0.22]{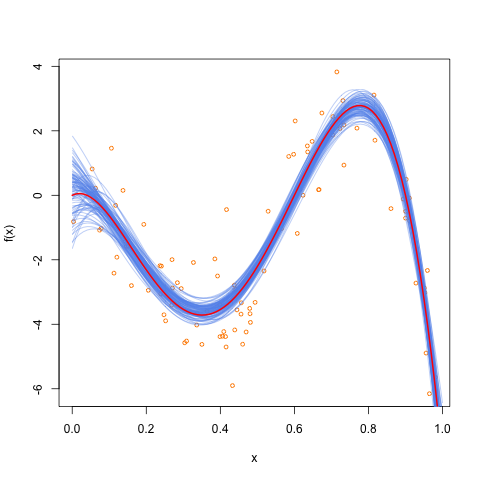}}\quad
\subfigure[SNR=2, $n$=200]{\includegraphics[scale=0.22]{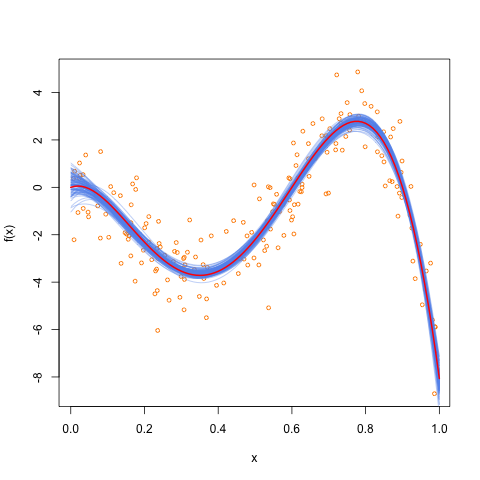}}\quad
\subfigure[SNR=2, $n$=500]{\includegraphics[scale=0.22]{Pred_n500SNR2fqff1.png}}}
\caption{Model fit for the selected order of smoothness in 100 datasets using the proposed method with mean function $\mu(x)=5x(5x-0.2)(0.4x-1.8)(3x-1.8)(2x-1.8)$} \label{fitf1}
\end{figure}

\begin{figure}[htbp]
\centering
\mbox{\subfigure[SNR=0.5, $n$=100]{\includegraphics[scale=0.22]{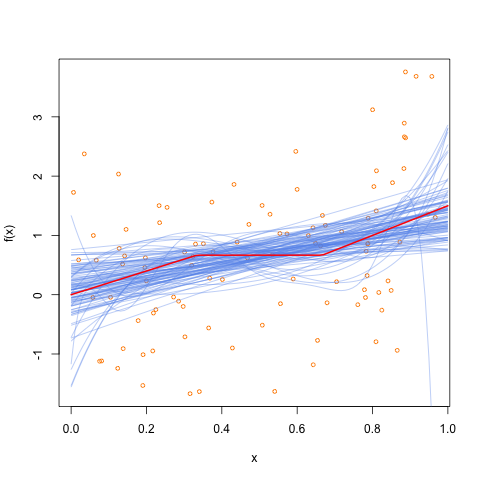}}\quad
\subfigure[SNR=0.5, $n$=200]{\includegraphics[scale=0.22]{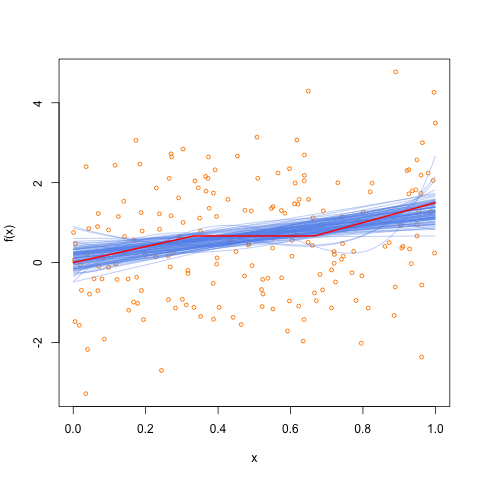}}\quad
\subfigure[SNR=0.5, $n$=500]{\includegraphics[scale=0.22]{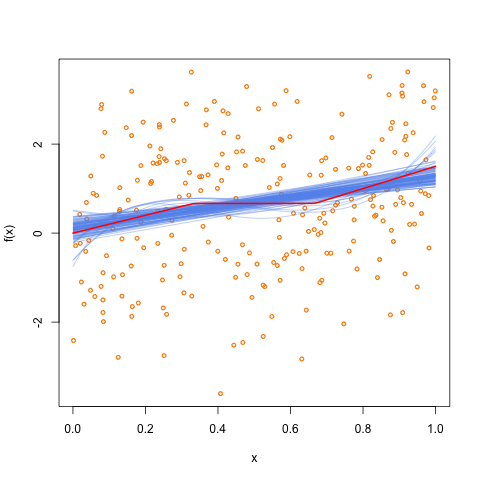}}}\\

\mbox{\subfigure[SNR=1, $n$=100]{\includegraphics[scale=0.22]{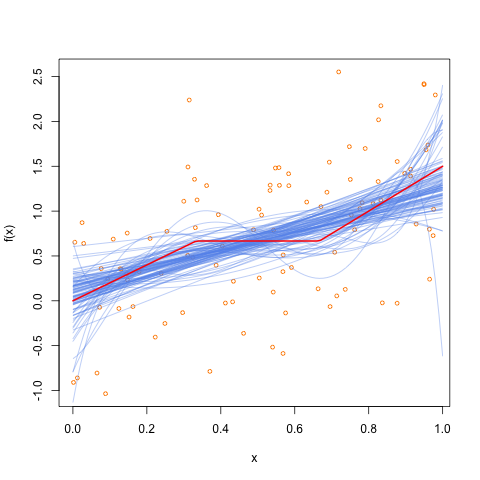}}\quad
\subfigure[SNR=1, $n$=200]{\includegraphics[scale=0.22]{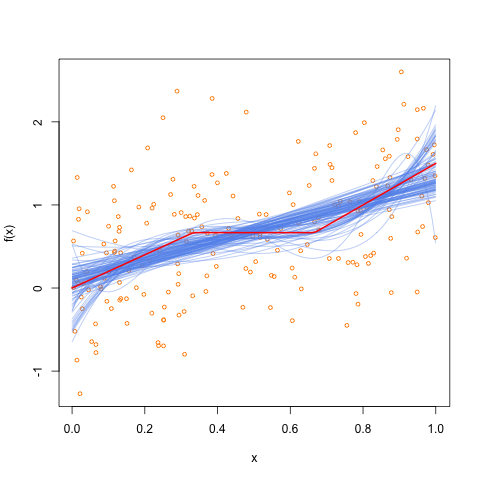}}\quad
\subfigure[SNR=1, $n$=500]{\includegraphics[scale=0.22]{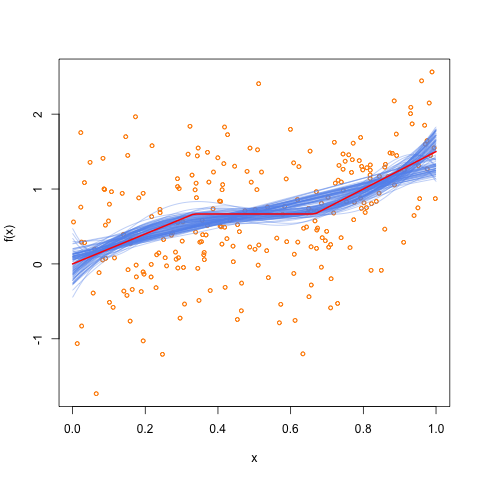}}}\\

\mbox{\subfigure[SNR=2, $n$=100]{\includegraphics[scale=0.22]{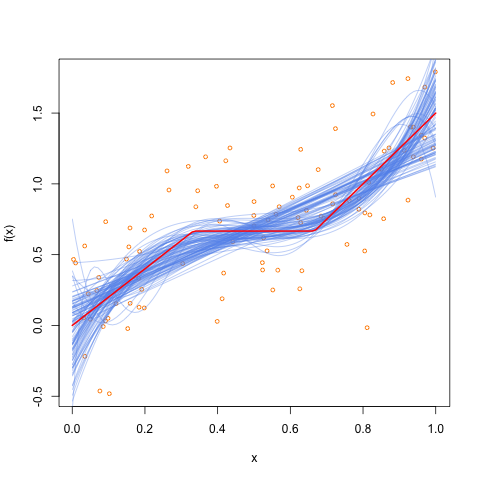}}\quad
\subfigure[SNR=2, $n$=200]{\includegraphics[scale=0.22]{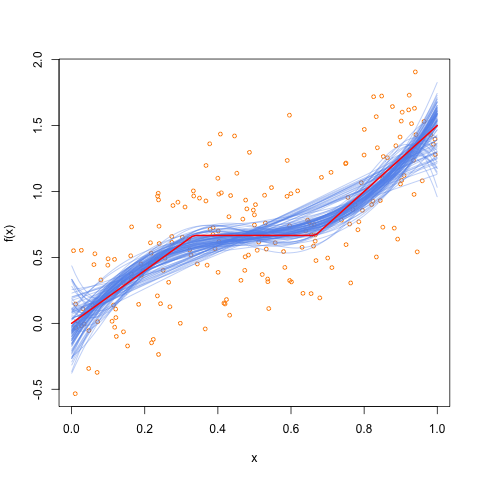}}\quad
\subfigure[SNR=2, $n$=500]{\includegraphics[scale=0.22]{Pred_n500SNR2fqff4.png}}}

\caption{Model fit for the selected order of smoothness in 100 datasets using the proposed method with mean function $\mu(x)=x \text{I}_{\lrb{x\in\lrsqb{-3,-1}}}-\text{I}_{\lrb{x\in\lrb{-1,1}}}+(x-2)\text{I}_{\lrb{x\in[1,3)}}$} \label{fitf4}
\end{figure}

\end{document}